%% file: main.tex
\documentclass{nature}
\usepackage[T1]{fontenc}
\usepackage{float}
\usepackage{graphicx}
\usepackage{pdflscape}
\usepackage{hyperref}
\usepackage{pgf,pgffor}
\usepackage{textcomp}
\usepackage{threeparttable} 
\usepackage{xcolor}
\usepackage{ulem}
\usepackage{lineno}
\usepackage{caption}
\usepackage{academicons}
\usepackage{pifont}
\usepackage{longtable}
\usepackage{amssymb}
\usepackage {amsmath}
\usepackage{cleveref}
\definecolor{orcidlogocol}{HTML}{A6CE39}
\usepackage[symbol]{footmisc}

\usepackage{subfigure}
\usepackage{academicons}
\usepackage{pifont}
\usepackage{longtable}
\usepackage{multirow}
\usepackage{supertabular}
\usepackage[resetlabels,labeled]{multibib}
\newcites{Main}{References}
\newcites{Method}{References}

\newcommand{\insight}{\textit{Insight}-HXMT}

\makeatletter

\let\saved@includegraphics\includegraphics
\AtBeginDocument{\let\includegraphics\saved@includegraphics}
\renewenvironment*{figure}{\@float{figure}}{\end@float}
\makeatother

\newcommand{\AFF}[1]{$^{\foreach\d[count=\ni]in{#1}{\ifnum\ni=1\ref{\d}\else,\ref{\d}\fi}}$}

\definecolor{dkblue}{RGB}{54, 86, 169}

\title{\insight ~and GECAM-C observations of the brightest-of-all-time
GRB 221009A}

\begin{document}
\maketitle

\centerline{\textbf{\insight \& GECAM collaboration}}

\input{authors.tex}

\input{affiliations.tex}

%\linenumbers

\begin{abstract}
GRB 221009A is the brightest gamma-ray burst ever detected since the discovery of this kind of energetic explosions.
However, an accurate measurement of the prompt emission properties of this burst is very challenging due to its exceptional brightness. With joint observations of \textit{Insight}-HXMT and GECAM-C, we made an unprecedentedly accurate measurement of the emission during the first $\sim$1800 s of GRB 221009A, including its precursor, main emission (ME, which dominates the burst in flux), flaring emission and early afterglow, in the hard X-ray to soft gamma-ray band from $\sim$ 10 keV to $\sim$ 6 MeV. Based on the GECAM-C unsaturated data of the ME, we measure a record-breaking isotropic equivalent energy ($E_{\rm iso}$) of $\bf \sim 1.5 \times 10^{55}$ erg, which is about eight times the total rest-mass energy of the Sun. The early afterglow data require a significant jet break between 650 s and 1100 s, most likely at $\sim950$ s from the afterglow starting time $T_{AG}$, which corresponds to a jet opening angle of $\sim {0.7^\circ} \ (\eta_\gamma n)^{1/8}$, where $n$ is the ambient medium density in units of $\rm cm^{-3}$ and $\eta_\gamma$ is the ratio between $\gamma$-ray energy and afterglow kinetic energy. The beaming-corrected total $\gamma$-ray energy $E_{\gamma}$ is $\sim 1.15 \times10^{51} \ (\eta_\gamma n)^{1/4}$ erg, which is typical for long GRBs. These results suggest that this GRB may have a special central engine, which could launch and collimate a very narrowly beamed jet with an ordinary energy budget, leading to exceptionally luminous gamma-ray radiation per unit solid angle. Alternatively, more GRBs might have such a narrow and bright beam, which are missed by an unfavorable viewing angle or have been detected without distance measurement.
\end{abstract}

\section*{Introduction}

Gamma-Ray Bursts (GRBs) are the most violent explosions in the universe\cite{2018pgrb.book.....Z}. 
They are traditionally divided into long GRBs (LGRBs) and short GRBs (SGRBs) based on their durations (separated at about 2 s)\cite{1993ApJ...413L.101K} and multi-variable criteria including empirical relationships (e.g. hardness ratio, spectral lag, minimum variability timescale)\cite{2002A&A...390...81A,2009ApJ...703.1696Z}.
LGRBs and SGRBs are believed to originate from massive star core collapse and binary neutron star vmergers\cite{2006ARA&A..44..507W,2017ApJ...848L..13A}%(woosley2006supernova; abbott2017gravitational),  
respectively. Both types of progenitors could form a central engine, a hyperaccreting black hole, or rapidly rotating, highly magnetized neutron star, which launches a pair of relativistic jets, one of which is beamed towards earth. The GRB prompt emission originates from an internal dissipation of energy, either through internal shocks\cite{1994ApJ...430L..93R} or magnetic dissipation\cite{2011ApJ...726...90Z}, which usually lasts for sub-seconds to thousands of seconds in duration, peaks from keV to MeV in spectrum, and carries most of radiation energy of a GRB. The afterglow\cite{1997ApJ...476..232M,1998ApJ...497L..17S}, originating from the deceleration of the jet by an ambient medium, lasts months to years and spans the full electromagnetic spectrum.

It is known that both LGRBs and SGRBs are collimated\cite{1999ApJ...519L..17S,1999ApJ...525..737R}, and LGRBs have a typical opening angle of $\sim5^\circ$ \cite{2001ApJ...562L..55F}. It is a long-standing mystery how a relativistic jet is launched and collimated and what the geometry and composition of the jet are. It is generally believed that GRB jets have an angular structure, with the core much brighter than its outer region\cite{1998ApJ...499..301M,2001ApJ...552...72D}. Therefore, the apparent brightness of a GRB depends on the intrinsic energetics of the jet and the viewing angle relative to the jet\cite{2002ApJ...571..876Z,2002MNRAS.332..945R}. 

Regardless of the geometry of the jet, a good indicator of the GRB energetics is the isotropic energy ($E_{\rm iso}$) of the prompt emission, which is estimated from the measured fluence and distance with the assumption of isotropic radiation from the jet. Observationally, $E_{\rm iso}$ spans many order of magnitudes among GRBs. During the 55 years of GRB observations, GRB 170817A has the lowest $E_{\rm iso}$ down to $10^{46}$ erg, while several most energetic bursts have $E_{\rm iso}$ up to $\sim 10^{54}$ erg. Some studies show that there is a cutoff around $10^{54}$ erg in the $E_{\rm iso}$ distribution \cite{2017ApJ...837..119A}.
However, it is still unknown whether GRBs could have a higher $E_{\rm iso}$ that can be up to $10^{55}$ erg. 

To estimate the true $\gamma$-ray energy ($E_\gamma$) of the GRB prompt emission, the opening angle of the jet should be measured. This could be derived from the jet break feature in the afterglow lightcurve\cite{1999ApJ...519L..17S,1999ApJ...525..737R}. However, it is usually hard to do so, especially if the jet break occurs at a very early time when observations are sparse and when the afterglow emission is contaminated by prompt emission of early flaring emission powered by the central engine.

\section*{Observations of GRB 221009A}
The \textit{Insight}-Hard X-ray Modulation Telescope (\textit{Insight}-HXMT) was triggered by GRB 221009A at 13:17:00.050 UT on October 9th, 2022 (denoted as $T_0$) during a routine ground search of bursts \cite{2022ATel15660....1T} with the CsI(Na) detectors of the High Energy X-ray telescope (denoted as HE/CsI hereafter) which has regularly detected GRBs \cite{2021MNRAS.508.3910C,2022ApJS..259...46S} and monitored the first GW electromagnetic counterpart \cite{Li:2017iup}.
From the observation timeline, \textit{Fermi}/GBM firstly reported the trigger of GRB 221009A \footnote{Note that the \insight ~trigger is 60 ms after the \textit{Fermi}/GBM trigger time.} in near real-time \cite{2022GCN.32636....1V}. The first accurate localization was provided by \textit{Swift}/BAT when it was triggered by the bright afterglow emission \cite{2022GCN.32632....1D}. Optical spectroscopy observations revealed the redshift of its host galaxy as $z=0.151$ \cite{2022GCN.32648....1D,2023arXiv230207891M}, corresponding to a luminosity distance of 745 Mpc \footnote{Assuming a flat $\Lambda$CDM cosmology with $H_0$=67.4 km $\rm s^{-1} \rm Mpc^{-1}$ and $\Omega_m$=0.315 \cite{2020A&A...641A...6P}.}. Remarkably, thousands of very high energy (VHE) photons up to 18 TeV were detected from this burst by LHAASO \cite{2022GCN.32677....1H}.
An impressive world-wide campaign was conducted and is still conducted to capture the multi-wavelength post-prompt emission signature of the GRB\cite{2023arXiv230111170F,2023arXiv230203829S,2023arXiv230207891M,2023arXiv230204388L}. We emphasize the observations of Global Advanced Rapid Network Devoted to Multi-Messenger Addicts (GRANDMA) that derived upper limit during the prompt phase (-1.1 in V band up to 390 s post GRB time) up to 20 days in multi-wavelength including the Low Energy X-ray telescope observations of \insight \cite{Kann:2023hlo}.
%}

Although GRB 221009A was detected by a fleet of high energy telescopes, the extreme brightness, however, caused problems to many instruments (e.g. \textit{Fermi}/GBM \cite{2022GCN.32642....1L},%\cite{2022GCN.32642....1L}, 
 \textit{Fermi}/LAT \cite{2022GCN.32637....1B,2022GCN.32658....1P,2022GCN.32760....1O,2022GCN.32916....1O}, %\cite{2022GCN.32637....1B,GCN.32658....1P,GCN.32760....1O,GCN.32916....1O}, 
 Konus-\textit{Wind} \cite{2022GCN.32668....1F},
%\cite{2022GCN.32668....1F}, \cite{2022ATel15660....1T,2022ATel15703....1G}, 
 \textit{AGILE}/MCAL+AC \cite{2022GCN.32650....1U} %\cite{2022GCN.32650....1U}, 
and \textit{INTEGRAL}-SPI/ACS \cite{2022GCN.32660....1G}), 
such as data saturation (i.e. data lost due to limited bandwidth), unrecoverable dead-time, pulse pileup, etc, which prevent them from making accurate temporal and spectral measurements \cite{2022GCN.32660....1G,2022GCN.32642....1L,2022GCN.32668....1F,2022GCN.32762....1K}. However, an accurate measurements of fluence, spectrum and energy are of critical importance to understand a GRB.
Here we report the accurate measurement of GRB 221009A with \insight ~and GECAM-C.

As shown in Panel A of Figure \ref{fig:NetLC}, the \insight ~HE/CsI triggered on the precursor emission (PE) which has a Fast Rise and Exponential Decay (FRED) shape lasting about 10\,s. After a $\sim170$\,s-long relatively quiescent stage, the main emission (ME) of the burst starts (the extremely bright part spans from $T_0$+220\,s to $T_0$+280\,s), which is followed by a flare episode (FE, $T_0$+350\,s to $T_0$+600\,s). After $T_0$+600\,s, the HE/CsI light curve (0.6-3 MeV) suddenly turns to a smooth evolution, and the photon index also becomes harder (about -1.62) and relatively stable (see Figure \ref{fig:FlareToAfterglow}). These features indicate that the HE/CsI data after $T_0$+600\,s is dominated by the early afterglow. 

There were two GECAM instruments in operation during this burst: GECAM-B and GECAM-C.
GECAM-B is a dedicated gamma-ray monitor microsatellite launched together with GECAM-A in December 2020, and GECAM-C is a similar gamma-ray monitor aboard the SATech-01 satellite launched in July 2022 \cite{zhangdlzhengwen:2023GECAMC}. Both GECAM-B and GECAM-C are capable of being triggered by bursts in-flight and can downlink the near real-time alerts; however, they did not send out trigger alerts for GRB 221009A for different reasons. 
For GECAM-B, this GRB was blocked by the Earth during the prompt emission phase. When the GRB became visible from $T_0$+1595 s, the flux was quite low and GECAM-B did not detect a very significant signal given the complicated background during that time.
For GECAM-C, the GRB was well detected, but the satellite was flying over the regions where the in-flight trigger was disabled (from the PE to ME) or trigger enabled but set not to send down trigger alerts (during the FE) (see {\bf Methods} for details).

Thanks to the dedicated design to observe extremely bright bursts for GECAM instruments and the special working mode for GECAM-C operating in the polar orbit (see {\bf Methods}), one of the GECAM-C/GRD detectors (i.e. GRD01) provided an accurate measurement of this brightest ME. Each GRD detctor is readout by two channels: high gain (HG) channel (about 6-300 keV) and low gain (LG) channel (about 0.4 - 6 MeV). These two channels are independent in terms of data processing, transition and dead-time calculation. For both channels, the dead-time is 4 $\mu$s for a normal event and $\geq 69~\mu$s for an overflow event (see {\bf Methods}). 

\begin{figure}
\centering
\includegraphics[width=0.6\textwidth]{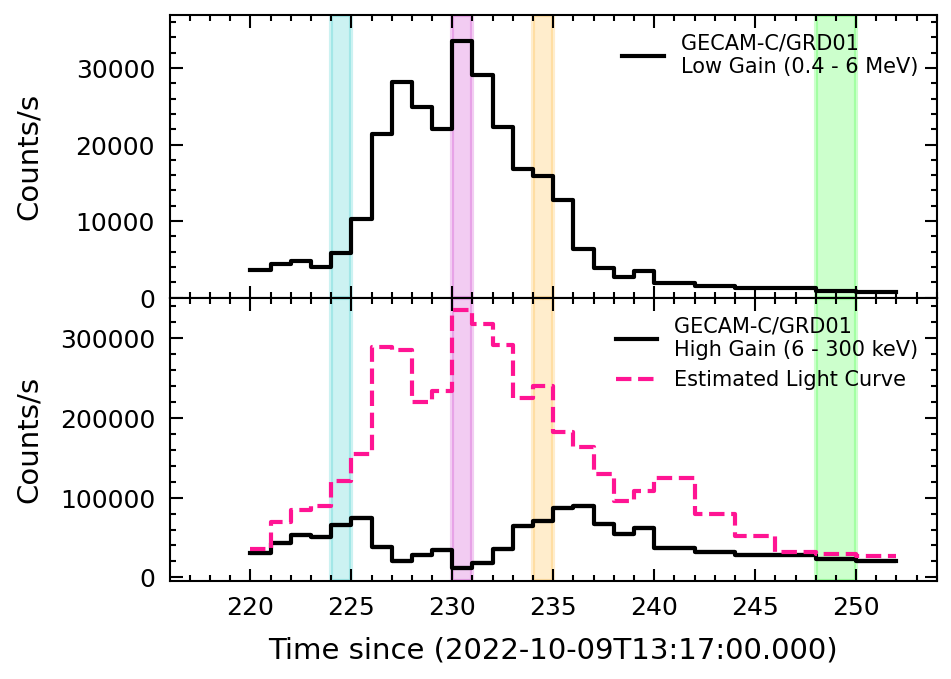}
\includegraphics[width=0.6\textwidth]{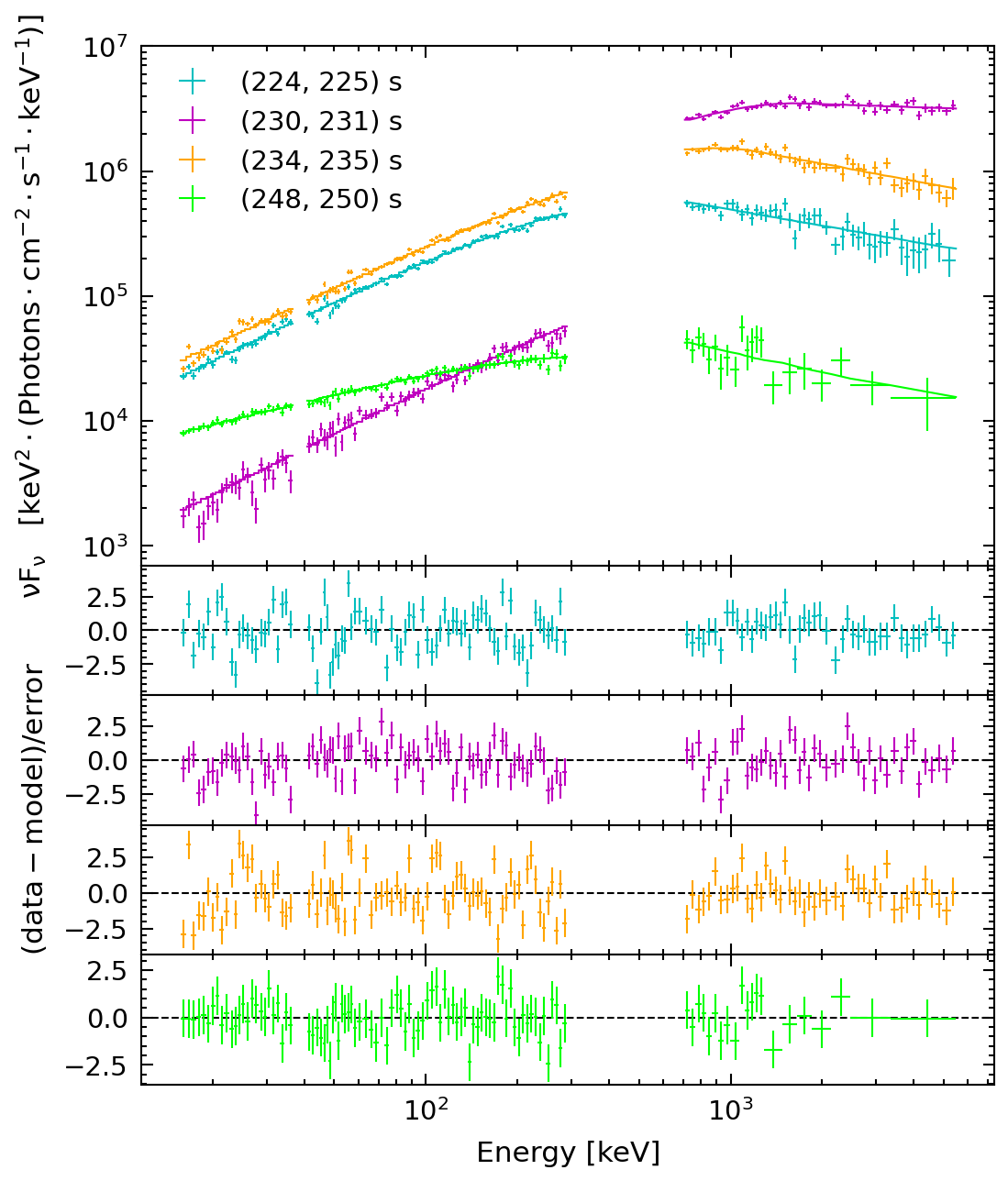}

\caption {The light curves and time-resolved spectral fitting of the brightest peak of GRB 221009A. {\bf Top panel:} The GECAM-C/GRD01 light curve of the brightest peak of this burst. The black solid lines on the light curves have been corrected for dead time for GRD01 LG and HG channels, while the dashed red line represents the estimated count rate from the incident GRB spectrum convolved with detector response. The difference between measured and estimated rates during the bright region is caused by the incorrect recording of deadtime and pulse pileup effects in HG. {\bf Bottom panel:} The spectral fit results for four time intervals.
A normalization factor has been applied between HG and LG to compensate for the extra counts loss in HG due to the incorrect recording of dead time and pulse pileup effect. }
\label{fig:Specfit}
\end{figure}

After a comprehensive inspection, we find both the HG and LG channels of GECAM-C/GRD01 are free of data saturation during the full burst including the brightest part of ME (see {\bf Methods}). During this burst, the maximum count rate in LG channel is about 30 k counts per second (see Top panel in Figure \ref{fig:Specfit}), which has negligible pulse pileup effect for the energy spectrum measurement (see {\bf Methods}). Also the dead-time is in the normal range and could be corrected with normal data analysis. On the other hand, the HG channel has much higher deposition count rate (up to 300 k counts per second, see Top panel in Figure \ref{fig:Specfit}) which causes some pulse pileup effect and incorrect measurement of dead-time for several time intervals with very high deposition rate (see {\bf Methods}); however, we use the LG channel data as the baseline to do the LG and HG joint analysis, thus these effects in the HG channel have very small impact on the joint spectral analysis (see {\bf Methods}).

\section*{Data Analysis}

\begin{figure}
\includegraphics[width=\textwidth]{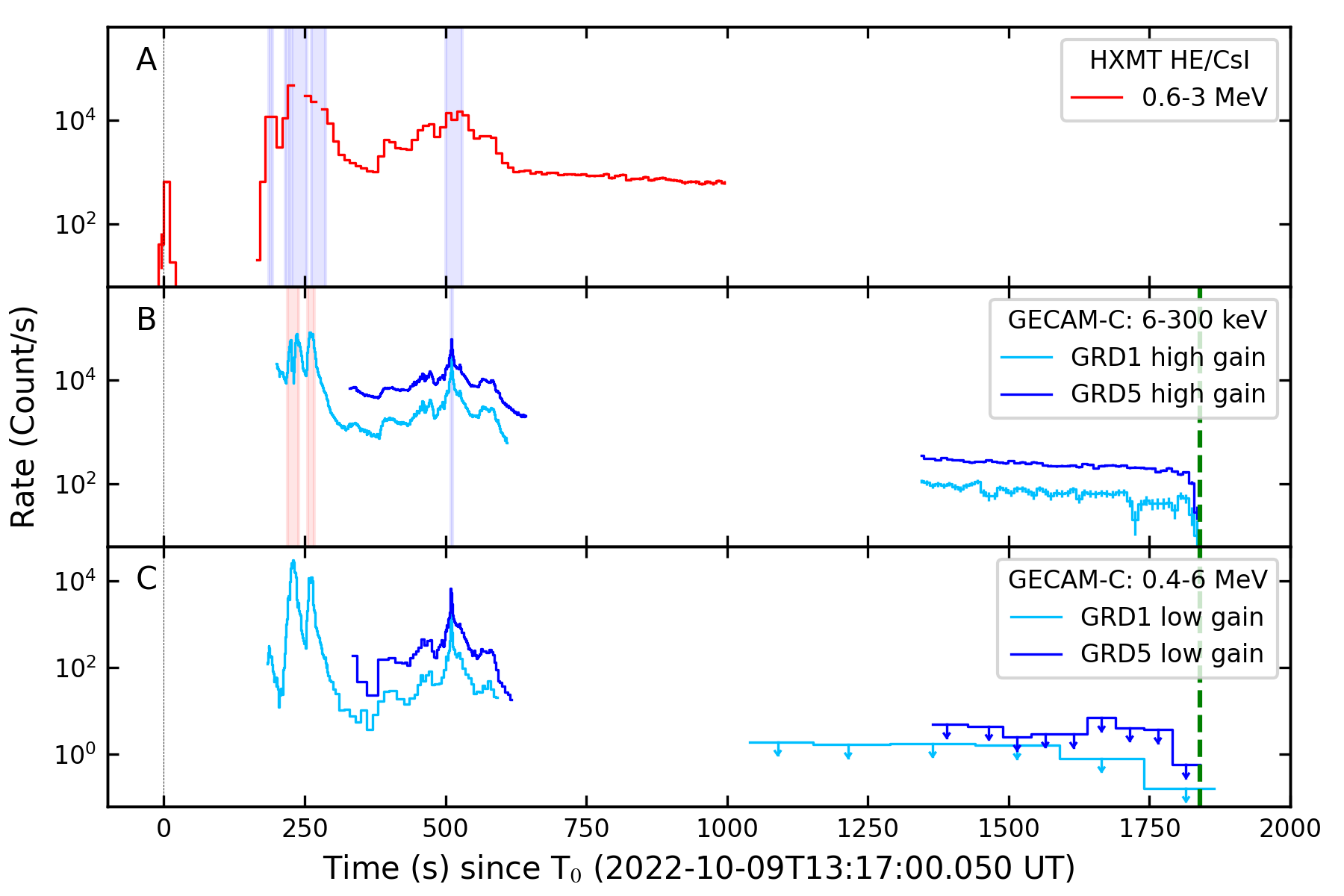}
\caption {Light curves of GRB 221009A with background subtracted. {\bf Panel A:} \insight ~HE/CsI light curve in 600--3000\,keV. The precursor is clearly detected without data saturation, but the ME and flare suffer data saturation (the purple rectangle regions). A clear trend of the early afterglow can be seen from $T_0$ + 600\,s to $T_0$ + 900\,s. {\bf Panel B:} GECAM-C/GRD light curves in High Gain (HG) channel. The GRD01 HG channel suffers from significant dead-time effects during the ME (the magenta rectangle regions). At the flare peak, data saturation briefly occurred in HG channel of GRD05 but not in GRD01. {\bf Panel C:} GECAM-C/GRD light curves in Low Gain (LG) channel. The LG channel of all GRDs (including GRD01 and GRD05) is free of data saturation or deadtime or pulse pileup issues.}
\label{fig:NetLC}
\end{figure}

With joint observations of \insight ~and GECAM-C, the full course of the prompt emission (from $T_0$ to $T_0$+300\,s), together with the flare episode (from $T_0$+350\,s to $T_0$+600\,s) and the early afterglow (from $T_0$+600\,s to $T_0$+900\,s and from $T_0$+1300\,s to $T_0$+1860\,s) are well measured in the hard X-ray to soft gamma-ray band (see Figure \ref{fig:NetLC}).

Since this burst lasts longer than usual GRBs detected by HE/CsI, we have applied the well-demonstrated \insight ~
background model technique \cite{2021ApJS..256...47Y}to HE/CsI data (see {\bf Methods}), which results in a net light curve for the burst (see Panel A in Figure \ref{fig:NetLC}).
The HE/CsI observation covers the precursor, main emission, flare and early afterglow before $T_0$+900\,s. 

The lightcurves of GECAM-C are shown in Panel B and C in Figure \ref{fig:NetLC}. Only GRD01 was turned on during the ME since the satellite was flying over a preset region, while all GRD detectors were on during the flare episode and the early afterglow. The background estimation is based on the recursive orbits. We have tested and evaluated this background estimation and find that it has negligible impact on the measurement of the ME and early afterglow (see {\bf Methods}).

To derive the energy flux and fluence\footnote{Flux and fluence are calculated in the energy range of 1-10000 keV in rest frame (i.e. bolometric), unless otherwise stated.}, a detailed time-resolved spectrum analysis is executed with different empirical spectral models, including Band function, cutoff power-law (CPL), power-law (PL) and blackbody (BB) (see {\bf Methods}). 
For \insight ~HE/CsI data in 0.6-3 MeV, the precursor is well fitted with a power law spectrum.

The most challenging part for GRB 221009A is the ME with ultra-high flux and the peak of the flare, which caused instrumental problems for many missions, including \insight ~HE/CsI (Panel A in Figure \ref{fig:NetLC}). Fortunately, GECAM-C worked in a special mode during that time with only one of 12 GRD detectors turned on, which significantly reduced the total data rate and avoided the data saturation (see {\bf Methods}). The GRD01 LG channel is free of data saturation or pulse pileup or dead-time problems, although the HG channel suffers issues of incorrect measurement of dead-time and slight pulse pileup during the most bright part of ME. However, both effects in HG channel do not distort the spectral shape significantly but underestimate the counts rate of HG (see {\bf Methods}). 
Thus we use the LG channel to measure the absolute normalization of spectrum, and employ the HG data to constrain spectrum shape in low energy band, and set a normalization factor between HG and LG to account for the underestimated counts rate in HG channel. See Figure \ref{fig:Specfit} for examples of the time-resolved spectral fitting for the burst.

\section*{Main Results on the Prompt Emission}

Based on the time-resolved spectral analysis results, we plot the flux light curve in Figure \ref{fig:FluxLC}. 
To make a complete light curve of the burst, the 20-200 keV flux is estimated from HE/CsI flux in 0.6-3 MeV for the precursor and early afterglow from $T_0$+600\,s to 900\,s (see {\bf Methods}). For the flux light curve in 20-200 keV, we find the $T_{90}$ is about 285\,s with the start time of about $T_0$+223.3 s.

The 1-s peak flux (from $T_0$+229.95\,s) is $\rm (1.72\pm 0.03) \times 10^{-2}\thinspace erg\cdot cm^{-2}\cdot s^{-1}$, corresponding to a peak luminosity of $\rm (1.14\pm 0.02) \times 10^{54} \thinspace erg\cdot s^{-1}$, while the 50-ms peak flux (from $T_0$+230.60\,s) is $\rm (2.62\pm 0.05) \times 10^{-2}\thinspace erg\cdot cm^{-2}\cdot s^{-1}$, corresponding to a peak luminosity of $\rm (1.74 \pm 0.03) \times 10^{54} \thinspace erg\cdot s^{-1}$. 

The bolometric fluence is $\rm ({2.36} \pm {0.14}) \times 10^{-5} \thinspace erg\cdot cm^{-2}$, $\rm(2.12\pm 0.02) \times 10^{-1} \thinspace erg\cdot cm^{-2} $, $\rm({6.03} \pm {0.30}) \times 10^{-3}\thinspace erg\cdot cm^{-2}$ for the precursor, ME and flare respectively, with a total fluence of $\rm(2.24\pm 0.02) \times 10^{-1} \thinspace erg\cdot cm^{-2} $. To compare with the previous record-breaking burst GRB 130427A which had a fluence of $2.5 \times 10^{-3} \thinspace \rm erg\cdot \rm cm^{-2}$ in 10-1000 keV, we calculate the 10-1000 keV fluence for GRB 221009A to be $ (1.20\pm 0.01) \times 10^{-1} \thinspace \rm erg\cdot \rm cm^{-2}$ for the full burst and $(4.15\pm 0.01) \times 10^{-3} \thinspace \rm erg\cdot \rm cm^{-2}$ for the flare. Note that even the flare has a larger fluence than that of GRB 130427A, and the full burst is $\sim$50 times brighter.

Regarding the spectral evolution behavior of the ME, the low energy index ($\alpha$) of Band function is almost always below the synchrotron death-line. The peak energy ($E_{\rm peak}$) shows flux-tracking behavior. We note that BB component is not required in the time-resolved spectrum throughout the burst. 

To calculate the averaged peak energy $E_{\rm peak}$, we implement a time-integrated spectral analysis, resulting in
$E_{\rm peak}=1247.4 \pm 91.2$ keV for the full burst. 
The ME is composed of two main peaks, denoted as Peak-1 and Peak-2 respectively (see top Panel in Figure \ref{fig:AmatiRelation}). With only Peak-1, its $E_{\rm iso}$ already makes it the record-breaking one among all detected GRBs. As shown in Figure \ref{fig:AmatiRelation}, the full burst, the two main peaks and the flare are all consistent with Amati relation \cite{amati2008measuring}.

\section*{Main Results on the afterglow}

\begin{figure}
\centering
\includegraphics[width=0.8\textwidth]{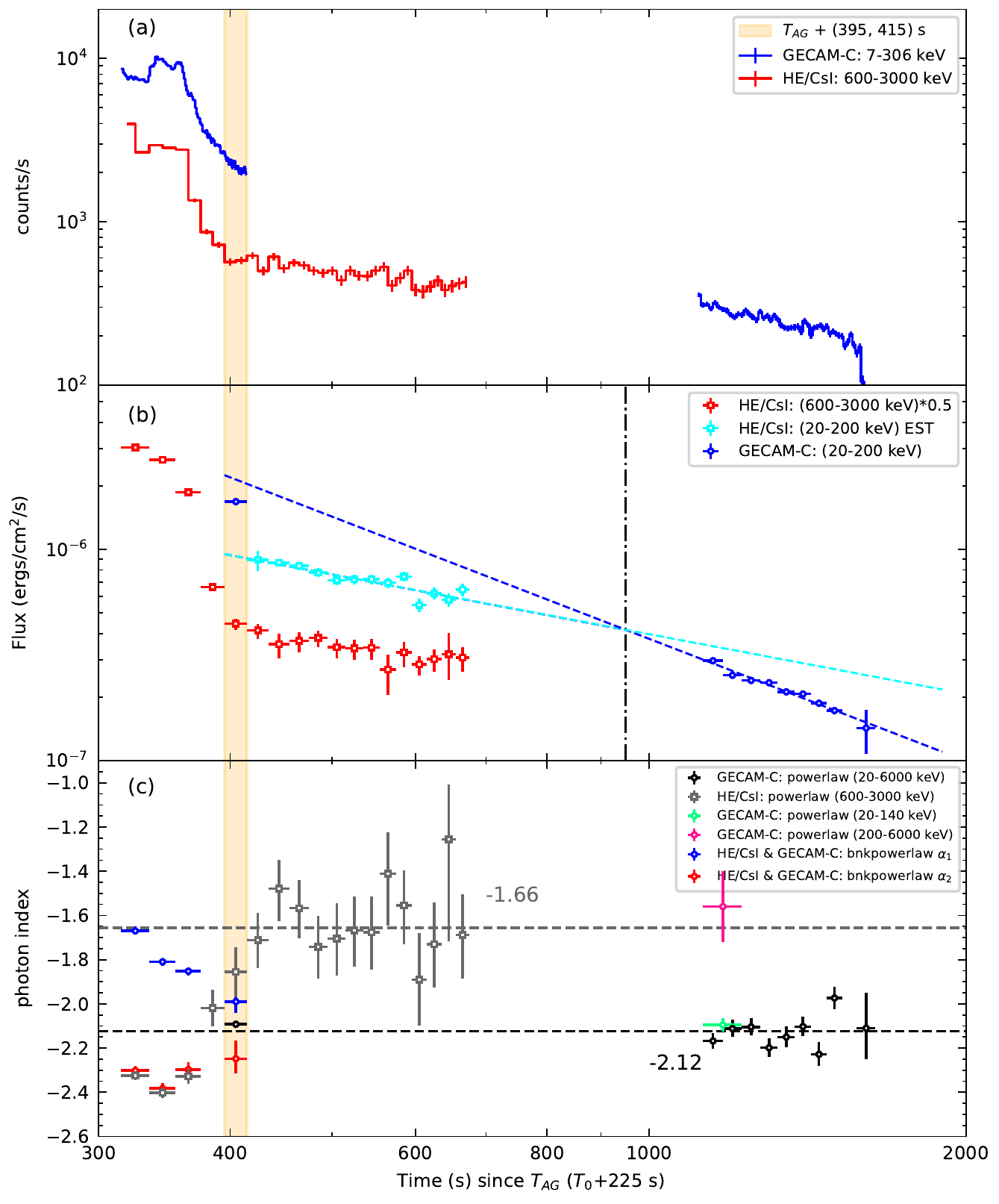}
\caption {Evolution of light curve (Panel a), energy flux (Panel b) and photon index (Panel c) from flare-dominated phase to afterglow-dominated phase of GRB 221009A. The dashed lines are fit results to the HE/CsI light curve from $T_0$+600\,s to $T_0$+900\,s and GECAM-C light curve from $T_0$+1300\,s to $T_0$+1860\,s. The jet break time is estimated from a joint fit with broken power-law to the estimated 20-200 keV flux from $T_0$+600\,s to $T_0$+900\,s and the measured 20-200 keV flux from $T_0$+1300\,s to $T_0$+1860\,s.}
\label{fig:FlareToAfterglow}
\end{figure}

We note that the flux light curve shows a large flare ($T_0$+350 s to $T_0$+600 s) on top of a decaying afterglow. 
%The evolution of early afterglow is also clearly detected.
From the light curve of \insight ~HE/CsI in 0.6-3 MeV, the afterglow should dominate the flux starting from $T_0$+600 s as the flare declines quickly in this high energy band, see Panel A in Figure \ref{fig:FlareToAfterglow}. The HE/CsI spectra in this time range could be well fitted with a single power law. The photon indices are consistent with an averaged value of -1.62 within error (Panel C in Figure \ref{fig:FlareToAfterglow}). 
 Since the precursor is quite weak and the afterglow is mostly contributed by the shells that power the most energetic prompt emission, we choose the approximate start time of the first main peak as the time zero of the afterglow ($T_{\rm AG}$ = $T_0$+225 s) to fit the afterglow light curve.
Fitting the HE/CsI flux light curve from $T_0$+600 s to $T_0$+900 s with a power law ($f(t) \propto t^{-\alpha}$) yields a slope $\alpha$= $-0.88 \pm 0.09$, which is consistent with the slope ($-0.88 \pm 0.14$) obtained in the direct fitting to the light curve. This is expected because the change in the energy response is rather small as the change in incident angle is also small during the scanning observation during this time (see {\bf Methods}). 

GECAM-C/GRD detectors made good observations of the afterglow from $T_0$+1350 s to $T_0$+1860 s, when the flare emission should be negligible compared to the afterglow emission even in the low energy band (i.e. 20-200 keV). The time-resolved spectral analysis of this time range also shows a consistent photon index (about -2.12) in the low energy band (20-200 keV). 
Only fitting the GECAM-C/GRD spectrum in the higher energy range (> 200 keV) gives a harder photon index (about -1.56$\pm 0.16$), which is generally consistent with the HE/CsI spectral index from $T_0$+600 s to $T_0$+900 s. This suggests that the afterglow spectral shape likely did not change significantly from $T_0$+600 s to $T_0$+1860 s (i.e. this early afterglow stage), and the photon index in the high energy band (> 200 keV) is harder than that in low energy band (< 200 keV).
Since the incident angle of the burst to GECAM-C detectors are constant (see {\bf Methods}), the detector response is the same for this time range, and we can directly fit the light curve instead of energy flux. This fit gives a slope of $-1.89 \pm 0.07$, which is well consistent with the slope of -$1.92 \pm 0.11$ of the energy flux decay within error (see {\bf Methods}).
We note that optical and soft X-ray data at later times shows a shallow decay slope (e.g. $\alpha_{opt}=-0.722\pm0.012$ \cite{Kann:2023hlo}, $\alpha_{x}\sim-1.5$ \cite{williams2023grb}), significantly shallower than the post-break slope found by GECAM-C. This could be related to the emergence of a broader structured jet wing surrounding the narrow core jet detected in the early time.

The prominent difference in the decay slope requires a break between these two segments of afterglow (see Panel (a) and (b) in Figure \ref{fig:FlareToAfterglow}). A significance test of this break with net light-curve results in $>5 \sigma$ (see {\bf Methods}). Since the spectral index did not change significantly as discussed above, this break is very unlikely caused by spectral evolution. Indeed, the slopes of about -0.9 and -1.9 strongly suggest that these two segments of afterglow are in the pre- and post-jet-break stage, respectively. Therefore the jet break should happen between them, i.e. $T_{\rm AG}+ 650$~s $\le t_{\rm j} \le T_{\rm AG}+ 1100$~s. 

\begin{figure}
\includegraphics[width=1\textwidth]{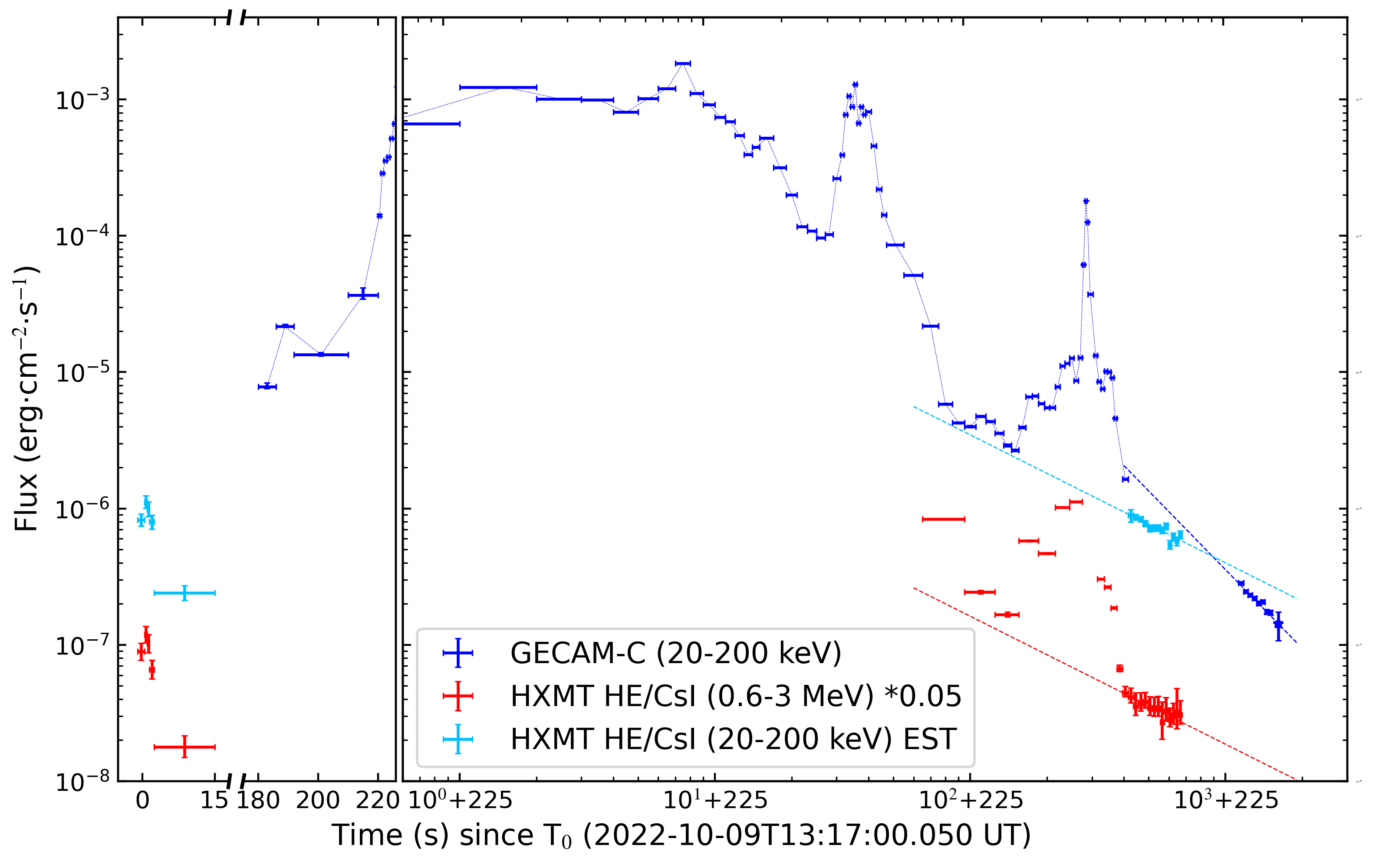}
\caption {Energy flux of GRB 221009A measured by \insight ~HE/CsI and GECAM-C. The 20-200 keV flux during precursor and early afterglow from $T_0$+600\,s to $T_0$+900\,s is estimated from HE/CsI measurement in 0.6-3 MeV according to the spectral shape of the afterglow (see Methods). Dashed-lines are fit results of the afterglow derived from the analysis in Figure \ref{fig:FlareToAfterglow}.}
\label{fig:FluxLC}
\end{figure}

To further estimate the jet break time, we assume that the afterglow spectral shape keeps the same from pre-break to post-break (which is also expected from the afterglow theory), and extrapolate the pre-jet-break HE/CsI flux in 0.6-3 MeV to estimate the HE/CsI flux in 20-200 keV as mentioned above, and then jointly fit with post-jet-break GECAM-C/GRD flux in 20-200 keV with a broken power law, yielding a break time $t_{\rm j}=T_{\rm AG}+950_{-50}^{+60} $ s. We note that the assumption on spectrum and extrapolation process could introduce additional errors to the jet break time. However we also note that the estimated 20-200 keV flux seems fairly reasonable as it generally follows the early afterglow trend underlying the flare episode (see Fig. \ref{fig:FluxLC}). The early jet break is consistent with that observed in the very high energy gamma-ray light curve of this burst by the LHAASO experiment (Cao et al. 2023, Science submitted).

Making use of
\begin{equation}
\theta_{\rm j} \simeq (0.063\, {\rm rad})\left({\frac{t_{\rm j}}{1\ {\rm day}}}\right)^{\frac{3}{8}} \left(\frac{1+z}{2}\right)^{-\frac{3}{8}}\left(\frac{{E_{\gamma,{\rm iso}}}}{10^{53}\ {\rm erg}}\right)^{-\frac{1}{8}}\times \left(\frac{{\eta}_{\gamma}}{0.2}\right)^{\frac{1}{8}} \left(\frac{n}{0.1\ {\rm cm}^{-3}}\right)^{\frac{1}{8}}.
\label{eq1}
\end{equation}
from Ref. \cite{2018pgrb.book.....Z} and plugging in the measured $t_{\rm j}$, $E_{\gamma,{\rm iso}}$ and $z$, we derive a jet opening angle of $\theta_{\rm j} \sim {0.71^\circ}^{+0.02^\circ}_{-0.01^\circ} (\eta_\gamma n)^{1/8}$, where $\eta_\gamma$ is the ratio between $\gamma$-ray and kinetic energies and $n$ is the ambient number density  in units of $\rm cm^{-3}$, which are left as free parameters.

\section*{Discussion and Conclusion}

According to the accurate measurements of \insight ~and GECAM-C, GRB 221009A has a record-breaking isotropic equivalent energy ($E_{\rm iso}$) of $\bf 1.5 \times 10^{55}$ erg, which is about eight times the rest-mass energy of the Sun. 
This huge energy was released in tens of seconds. The isotropic equivalent peak luminosity ($L_{\rm iso}$) is $\rm \sim1.7 \times 10^{54} \thinspace erg\cdot s^{-1}$ (in the timescale of 50 ms), which is among the highest $L_{\rm iso}$ GRBs. Standard GRB central engine models invoke a stellar-mass black hole, which should have an accretion rate of $\ll 1 {M}_\odot \ {\rm s}^{-1}$ because of the limitation of the total available mass from the progenitor star. In order to accommodate the huge isotropic energy and luminosity, significant jet collimation is needed.

Our derived opening angle $\theta_{\rm j} \sim 0.7^{\circ}$ is the smallest jet opening angle ever detected. One can use it to correct for the true $\gamma$-ray energy using the beaming factor $f_{\rm b} = 1-\cos\theta_{\rm j} \simeq \theta_{\rm j}^2/2 \sim 7.5\times 10^{-5}$. This gives $E_\gamma = f_{\rm b} E_{\rm \gamma,iso} \sim 1.15 \times 10^{51}$ erg. It has been known that long GRBs have a quasi-standard energy reservoir with bursts with high isotropic energies ($E_{\rm \gamma,iso}$) having smaller jet opening angles \cite{2001ApJ...562L..55F}, even though there is significant scatter to the correlation\cite{2018ApJ...859..160W}. Our result is broadly consistent with such a correlation, with the beaming corrected energy consistent with the distribution of $E_\gamma$. The small $\theta_{\rm j}$ nicely explains the enormously large $E_{\rm \gamma,iso}$ and the very small chance of detecting such a burst. So far there are about $10^4$ GRBs detected in the universe. The true number is about $10^4 / \bar f_b$, where $\bar f_b$ is the average beaming correction factor of all GRBs. If all GRBs have such a narrow bright core, the expected $E_{\rm \gamma,iso} \gtrsim 10^{55}$ erg GRBs that should have been detected is $\sim 10^4 (f_b/ \bar f_b)$. Given the uncertainty of $\bar f_b$ (which depends on jet structure), this number should be $\sim (10-100)$. One would expect most of these $E_{\rm \gamma,iso} \gtrsim 10^{55}$ erg GRBs located at higher redshifts ($z> 0.15$) whose redshifts (and hence, $E_{\rm \gamma,iso}$) were not measured. Alternatively, it is possible that not all GRBs have such a narrow bright core. In any case, the fact that such an event was detected at $z \sim 0.15$ has a small chance possibility, but may be accommodated by small-number statistics.

Even though the beaming corrected energy $E_\gamma \sim 1.15 \times 10^{51}$ erg does not demand an exceptional central engine in terms of energy budget, the extremely narrow $\sim 0.7^\circ$ jet inferred from the jet break indeed calls for a special engine that can collimate the jet to such a high degree. One possibility is that a highly magnetized central engine launched the jet through the Blandford-Znajek effect\cite{1977MNRAS.179..433B,2000PhR...325...83L}. The engine may have a dense, thick accretion torus or the progenitor may have a dense inner stellar envelope of a collapsing star, which can collimate the jet to a higher degree than normal GRBs. Alternatively, all GRB progenitors may be able to collimate such narrow jets, but for the majority of bursts, the Earth observer may have missed this narrow beam and only see the broader region of the jet, or have detected them but unable to identify the ultra-high $E_{\rm \gamma,iso}$ due to the absent distance measurement. Future detailed modeling of this GRB and the entire GRB population is needed to test both possibilities.

\noindent \textbf{Figures}

\begin{figure}
\centering
\includegraphics[width=0.8\textwidth]{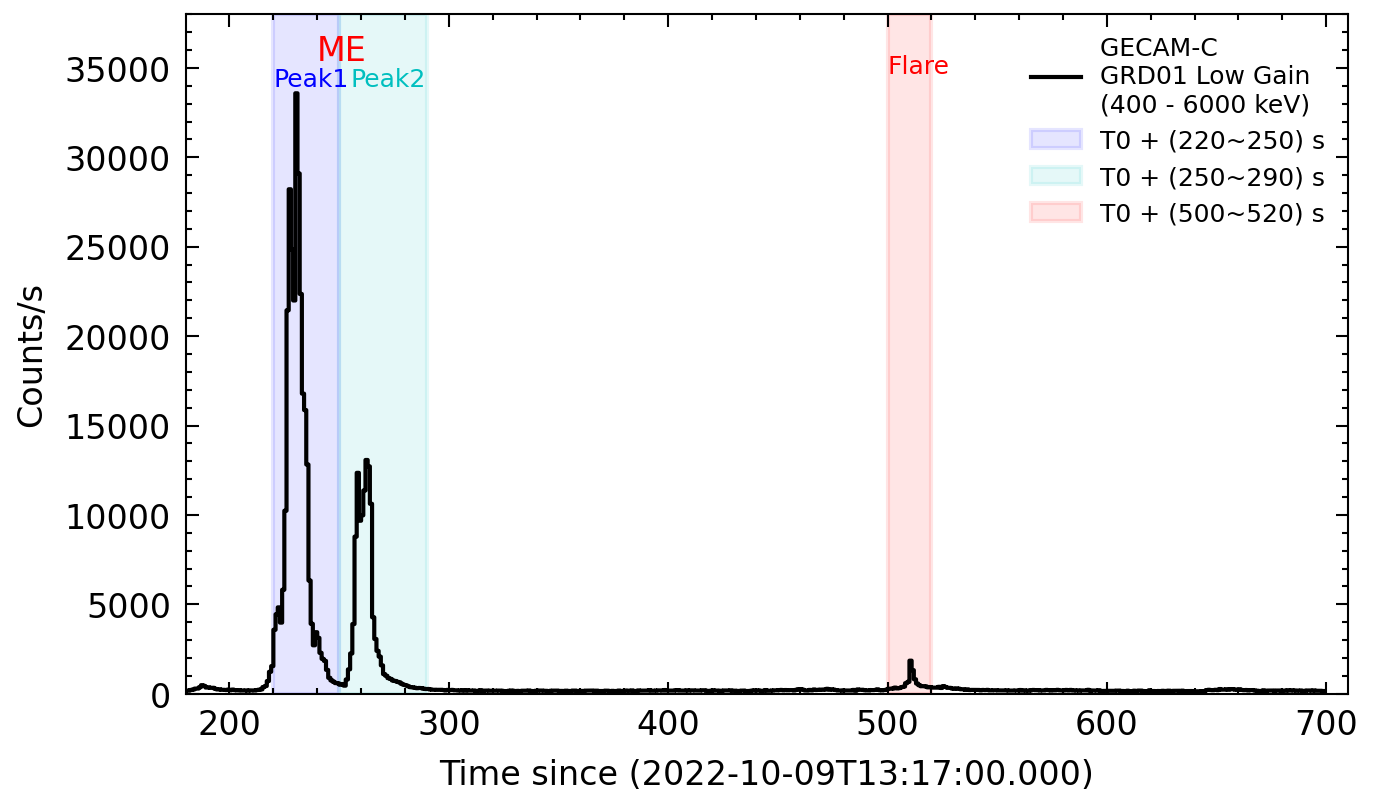}
\includegraphics[width=0.8\textwidth]{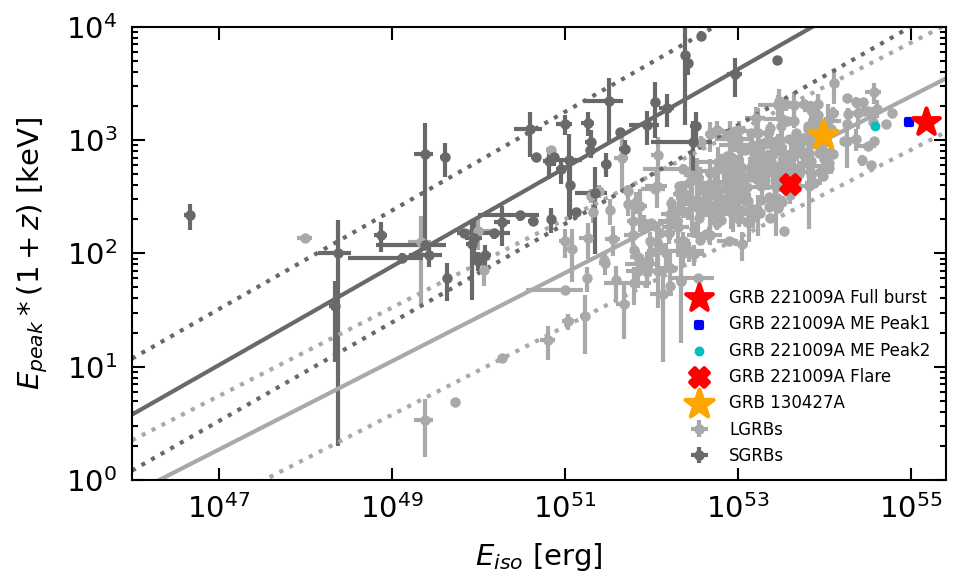}
\caption {{\bf Top Panel:}  GECAM-C/GRD01 Low Gain light curve of GRB 221009A. {\bf Bottom Panel:} Peak energy $E_{\rm p}$ and isotropic energy $E_{\rm iso}$ for GRB 221009A. The results for the two peaks in ME (Peak-1 and Peak-2) and the bright part of flare are also shown.}
\label{fig:AmatiRelation}
\end{figure}

\clearpage
\renewcommand{\refname}{References}
\bibliographystyle{naturemag}
\bibliography{maintex}
\clearpage

\subsubsection*{Data availability} 
The data that support the plots within this paper and other findings of this study are available from the \textit{Insight}-HXMT and GECAM website (http://www.hxmt.cn/ and http://gecam.ihep.ac.cn/).

\subsubsection*{Code availability}
The \textit{Insight}-HXMT and GECAM data reduction were performed using software available from the \textit{Insight}-HXMT and GECAM website (http://www.hxmt.cn and http://gecam.ihep.ac.cn/). The model fitting of spectra was completed with XSPEC, which is available from the HEASARC website (https://heasarc.gsfc.nasa.gov/).

\begin{addendum}

\item[Acknowledgements] 

This work used data from the \textit{Insight}-HXMT mission and GECAM (Huairou-1) mission. The \textit{Insight}-HXMT mission is a project funded by the China National Space Administration (CNSA) and the Chinese Academy of Sciences (CAS). the GECAM (Huairou-1) mission is funded by the Strategic Priority Research Program on Space Science (XDA15360000) of the Chinese Academy of Sciences (CAS). We gratefully acknowledge the support from the National Program on Key Research and Development Project (Grant No.2021YFA0718500) from the Minister of Science and Technology of China (MOST). The authors thank supports from the National Natural Science Foundation of China under Grants 
(12273042, %% Shaolin Xiong
12273043, %%  Xiaobo Li
U2038106, %% Chengkui Li
12021003, %% He Gao
12103055, %% Bing Li
U1938201, %%Bing Li
12173038) %% Xinqiao Li
and the Science Foundation of Hebei Normal University (No.L2023B11). 
This work was partially supported by International Partnership Program of Chinese Academy of Sciences (Grant No.113111KYSB20190020).
J.-R. Mao is supported by the National Natural Science Foundation of China 11673062 and the Oversea Talent Program of Yunnan Province. 
D.~A.~Kann acknowledges the support by the State of Hessen within the Research Cluster ELEMENTS (Project ID 500/10.006).
MWC is supported by the National Science Foundation with grant numbers PHY-2010970 and OAC-2117997.

\item[Author Contributions] S.-L.X. led this work including data analysis, discussion and manuscript as the PI of GECAM mission and the team leader of \textit{Insight}-HXMT GRB research. 
S.-N.Z. led the \textit{Insight}-HXMT observation and research as the PI of \textit{Insight}-HXMT. 
B.Z. led the theory interpretation. 
Y.-Q.Z., C.Z., J.-C.L., W.-C.X., Y.Y., C.-W.W., W.-J.T., S.-L.X., C.-K.L., M.-Y.G., J.-Y.L., X.-B.L., C.-Z.L. are the main contributor to the data analysis.
B.Z., Z.-G.D., X.-Y.W., H.G., S.-N.Z., S.-L.X., Z.Z., Y.-W. Y., J.-R.M., L.-J.W., B.L. participated in the theory interpretation and discussion.
S.-L.X., S.-N.Z., B.Z., C.-K.L., Y.-Q.Z., C.Z., J.-C.L., W.-C.X., C.-W.W., Y.Y., C.C., S.X., J.-Y.L., W.-J.T., C.-Z.L., F.-J.L. participated in the writing and editing of the paper.
All authors participated in the discussion and proofreading of the manuscript, or contributed to the development, operation or GRB research of \textit{Insight}-HXMT or GECAM, which is important for this work.

 \item[Competing Interests] The authors declare no
competing financial interests.

\item[Correspondence] Requests for materials should be addressed to the following: \\
Shao-Lin Xiong (E-mail: xiongsl@ihep.ac.cn), Shuang-Nan Zhang (E-mail: zhangsn@ihep.ac.cn), Bing Zhang (E-mail: bing.zhang@unlv.edu).

\end{addendum}

\clearpage

\clearpage

\clearpage

\section*{Methods}

\subsubsection*{\insight ~and GECAM observations}

The \textit{Insight}-Hard X-ray Modulation Telescope (\textit{Insight}-HXMT) is China's first X-ray astronomy satellite \citeMethod{2020SCPMA..6349502Z,2020JHEAp..27...64L} which was launched on June 15th, 2017 at an altitude of 550 km and an inclination of 43 degree. It features a broad X-ray range (1-250 keV) and consists of three telescopes: the High Energy X-ray telescope (HE) which uses 18 NaI(Tl)/CsI(Na) phoswich scintillation detectors for 20-250 keV \cite{2020SCPMA..6349503L}, the Medium Energy X-ray telescope which uses 1728 Si-PIN detectors for 5-30 keV \cite{2020SCPMA..6349504C}, and the Low Energy X-ray telescope which uses 96 Swept Charge Device (SCD) detectors for 1-15 keV \cite{2020SCPMA..6349505C}. All three telescopes use slat collimators to confine their field of views. The CsI scintillation detectors can effectively monitor the all-sky in gamma-ray band. 
Detailed studies show \textit{Insight}-HXMT works in a good status \cite{2023arXiv230203859L,2023arXiv230210714L,2023arXiv230214459L}.
 
When GRB 221009A occurred, the HE/CsI detectors of \textit{Insight}-HXMT operated in the low voltage mode (PMT voltage lower than normal value) with the energy range of about 200-3000 keV (deposited energy). Only gamma-rays with energy greater than about 200 keV can penetrate the spacecraft and leave signals in the HE/CsI detectors installed inside of the telescope. 

\begin{figure}
\includegraphics[width=\textwidth]{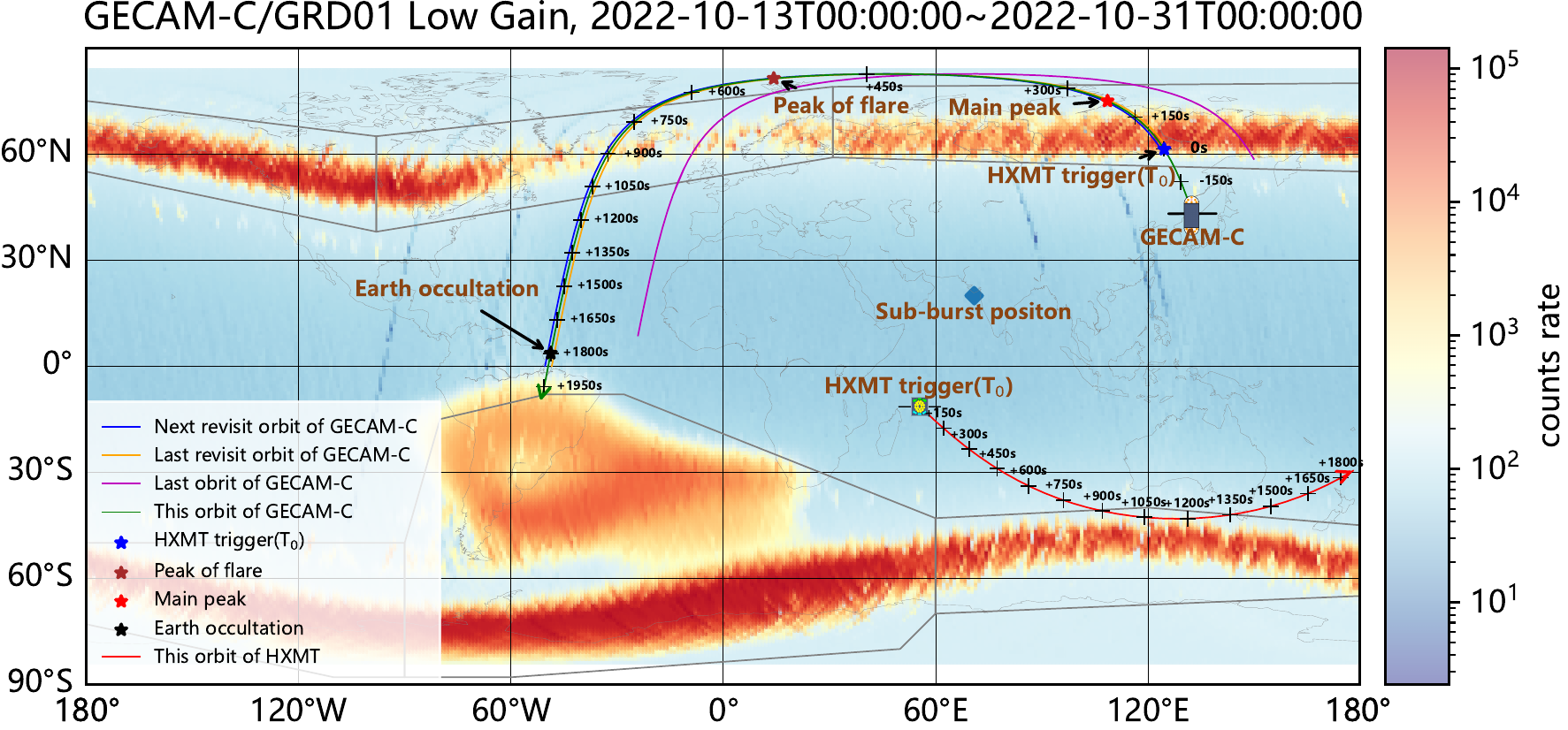}
\caption {Satellite track of \insight ~and GECAM during the GRB 221009A observations.}
\label{fig:SatelliteTrack}
\end{figure}

\begin{figure}
\includegraphics[width=\textwidth]{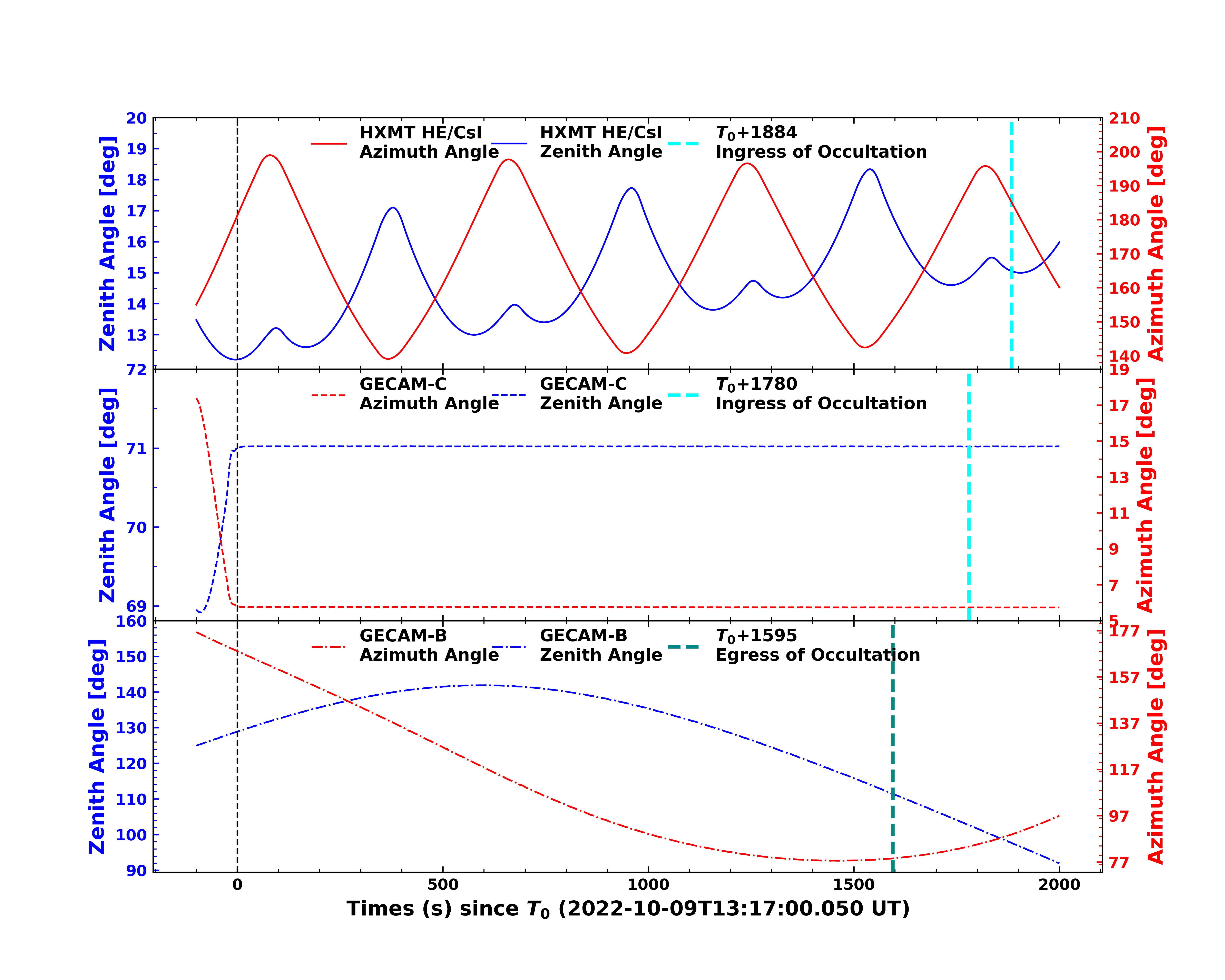}
\caption {Incident angles of \insight ~and GECAM during the GRB 221009A observations.}
\label{fig:IncidentAngle}
\end{figure}

 From the precursor to the ME and the flare, as well as the early afterglow phase, GRB 221009A was visible to \insight \ until it entered the Earth's shadow from $T_0$+1884 s. However, after about $T_0$+900 seconds, the \insight \ entered the high-latitude region of the Earth, and the estimation of the background level is of high uncertainty (see Figure \ref{fig:SatelliteTrack}).
During the full burst of GRB 221009A, \insight ~was making scan observations to a small sky area, and its pointing direction was moving (see Figure \ref{fig:IncidentAngle}), so the incident angle of GRB 221009A to the CsI detectors was changing. For the time-resolved spectral analysis, we make the CsI energy response matrix according to the incident angle averaged in the time interval.

The HE/CsI detector was triggered by GRB 221009A at $T_0$ during a routine ground search of bursts \cite{2022ATel15660method....1T}. 
%(cite hxmt atel \#15660). 
The trigger is due to the precursor lasting about 10 s. After a relative quiescent period of about 170\,s, the extremely bright ME ($T_0$+220\,s to $T_0$+280\,s) comes, which is followed by a flare episode around $T_0$+350\,s to $T_0$+600\,s. 

The ME and the flare peak are so bright that other detectors (Anti-Coincidence shield Detectors -- ACDs, Particle Monitors -- PMs) of HE as well as the medium energy telescope and low energy telescope also recorded the signal though their detection efficiency to gamma-rays is very low. Data saturation occurred in HE/CsI during some time period with ultra high count rate (See \cite{2020JHEAp..26...58X} for details), while PMs are free of such saturation because of the relatively low count rates due to the very small detector areas. Especially, the high rates of PMs triggered the threshold of turning off all detectors HE, medium energy telescope and low energy telescope from about $T_0$+230\,s to $T_0$+250\,s and $T_0$+268\,s to $T_0$+278\,s, just as \insight ~was entering the working Mode of SAA region. 
The low energy telescope of \insight also made afterglow observations of GRB 221009A, which is reported elsewhere \cite{Kann:2023hlo_method,2022ATel15703....1G}.

The GECAM mission was originally composed of two microsatellites (GECAM-A and GECAM-B) launched on Dec 10th 2020, and is extended with the operation of the GECAM-C (also called HEBS) \cite{zhangdl:2023GECAMC} which was launched onboard the SATech-01 satellite on July 27th, 2022. With the same design, each of GECAM-A and GECAM-B carries 25 gamma-ray detectors (GRD) and 8 charged particle detectors (CPD). GECAM-C makes use of the flight-spare detectors and electronics of GECAM-A and GECAM-B, and has the similar design and share the same science operation and data analysis pipeline as GECAM-A and GECAM-B. Since only GECAM-C data have been used in this work, in the following we focus on the GECAM-C. 

GECAM-C is equpped with 12 GRD detectors, including 6 LaBr3-based detectors (GRD\#01, 03, 05, 07, 09, 11) and 6 NaI-based detctors (GRD\# 02, 04, 06, 08, 10, 12.) \cite{zhangdl:2023GECAMC}. There are two detector domes placed at the top and bottom side of the SATech-01 satellite respectively. Each detector dome is configured with 6 GRDs with different pointing directions. In this way, the GRD detectors can monitor the whole sky unocculted by the Earth.

Each GRD detector is about 3 inches in diameter and has a geometric area of about 45 cm$^2$. All GRDs (except GRD06 and GRD12) are readout by two channels: HG channel (about 6-300 keV) and LG channel (about 0.4 - 6 MeV).

From $T_0$-68\,s to $T_0$+329\,s (including the precursor and ME) and from $T_0$+678\,s to $T_0$+1104\,s, GECAM-C was flying over the preset regions which cover high particle flux area. During this regions, GECAM-C was commanded to automatically work in a special mode for which only one gamma-ray detector (i.e. GRD01) and one charged particle detector (i.e. CPD02) are turned on, and the in-flight trigger is disabled. From $T_0$+329\,s to $T_0$+678\,s, GECAM-C was out of these regions thus it was switched to the normal working mode for which all detectors were turned on and in-flight trigger was enabled. Indeed it was triggered multiple times during the flare episode of this GRB, however, all in-flight triggers were set not to send down real-time alerts during this so-called high latitude area. From $T_0$-68\,s to $T_0$+170\,s and from $T_0$+645\,s to $T_0$+1350\,s, the background was very high or contaminated the burst signal, such that we cannot see the burst signal (e.g., the precursor in light curve). Overall, GECAM-C could effectively detect GRB 221009A (including ME and FE and early afterglow) from $T_0$+170\,s to $T_0$+645\,s and from $T_0$+1350\,s to $T_0$+1860\,s, when GRB 221009A was blocked by the Earth (i.e. Earth occultation).

The energy gain and detector response have been comprehensively calibrated for \insight~HE/CsI \cite{2020JHEAp..27....1L,2023arXiv230203859L} and GECAM-C/GRD \cite{zheng:2023ground_calibration, Zhangyq:2023cross_calibration}. A cross-calibration between HE/CsI and GECAM-C/GRD LG are also implemented for this burst (see Fig. \ref{fig:HEBS_CsI_high_index_calibration}). Especially, the high count rate performances of GECAM instruments have been studied in detail (Liu et al., to submit). All these work ensure us to properly understand the data and derive reliable results.

\subsubsection*{Calibration of \insight ~HE/CsI and GECAM/GRD}
As an active shield detector, \insight ~HE/CsI is designed to suppress the background of HE/NaI and also serves as an all-sky gamma-ray burst monitor \cite{2020JHEAp..27...64L}.
In the normal mode, the detected energy range of HE/CsI is about 40--800\,keV and in the low voltage mode it becomes 0.2--3\,MeV by reducing the high voltages of the photomultiplier tubes.

The gain of \insight ~HE/CsI shows a continuous increasing-trend after launch \cite{2023arXiv230203859L}. Routinely we calibrate the gain each month with six background characteristic lines from blank sky observations for the low voltage mode to guarantee the credibility of the E-C relationship of HE/CsI. The residuals for the gain calibration are less than 2\% for all HE/CsI detectors. At the same time, the energy response file calibration is also implemented by the simultaneous observations of typical bright GRBs with other GRB telescopes \cite{2020JHEAp..27....1L}. Benefit from the regular good calibration of HE/CsI, the data of GRB 221009A observed by HE/CsI can be correctly analyzed as HE/CsI was in the low voltage mode during the burst. 

\begin{figure}
\includegraphics[width=\textwidth]{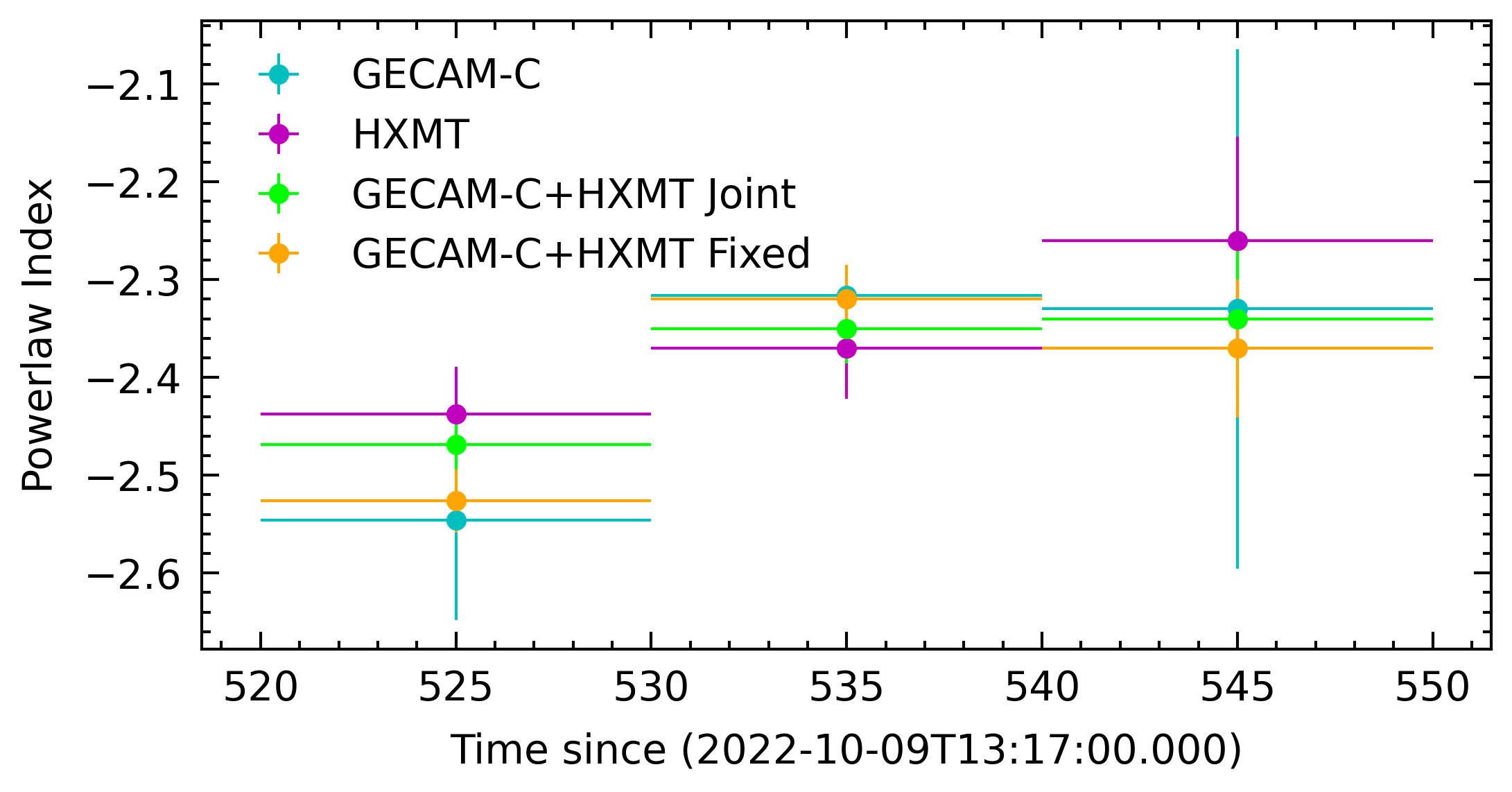}
\caption {Calibration of \insight ~and GECAM-C high-energy power-law spectral indices.}
\label{fig:HEBS_CsI_high_index_calibration}
\end{figure}

The energy gain calibration of GECAM-C/GRD detectors was implemented before the launch \cite{zheng:2023ground_calibration}.  The cross calibration of the GECAM-C/GRD detector response has also been made with Fermi/GBM and Swift/BAT \cite{Zhangyq:2023cross_calibration}. We also did a cross calibration between HE/CsI and GECAM-C/GRD for this burst, as shown in Figure \ref{fig:HEBS_CsI_high_index_calibration}. All these calibrations show the detector response of HE/CsI and GECAM-C GRD are correct and applicable for this burst analysis.

\subsubsection*{\insight~HE/CsI background modelling}

The background of \insight~HE/CsI is estimated by a parametric model, in which some main physical processes between the space environment and the satellite are simulated to reproduce the background. It is known that the charged particles in space environment, consisting largely of electrons and protons, collide with the satellite and generate the prompt background (charged particles deposit its energy directly in the scintillators) and the delayed background (the satellite materials activated by the charged particles decay, which is detected by the scintillators) \cite{2020JHEAp..27...14L}. The parametric model takes the flux of the charged particle as input data, and outputs the background light curves separately in different energy bands. This method has been proved effective and reliable on estimating the background of \insight \cite{2021ApJS..256...47Y2}. 
In the parametric model, the HE/CsI background is separated into three components:
\begin{equation}\label{equation:3components}
   B(t) = P(t)+D(t)+C(t),
\end{equation}
where $B(t)$ is the total background of HE/CsI, $P(t)$ represents the prompt background, $D(t)$ represents the majority of delayed background and $C(t)$ is the constant component representing those delayed background whose decay constant is so large that its intensity almost remain the same in the studied time period. The total count rate of 12 ACDs on the lateral side of \insight~ is chosen as the model input to represent the charged particle flux. The prompt components is set to be a polynomial function of ACD count rate:
\begin{equation}
  P(t) = a_{1} \cdot f^{3}(t) + a_{2} \cdot f^{2}(t) + a_{3} \cdot f(t),
\end{equation}
where $f(t)$ is total count rate of 12 ACDs on the lateral side of \insight, $a_{1}$, $a_{2}$ and $a_{3}$ are the polynomial coefficients. The constant component is represented by a constant parameter:
\begin{equation}
  C(t) = a_{4}.
\end{equation}
The delayed component is the decay signal of the activated materials in the satellite, induced by the charged particles in space environment. We simply assume that there are two kinds of radioisotopes with different decay constants causing delayed backgrounds $\lambda$ in each energy band, and the production rate of every kind of radioisotope is generally proportional to the intensity of the prompt component $P(t)$, with a slight modulation by the altitude of the satellite. Thus, expression of the delayed component of the background is:
\begin{equation}
  D(t)=\sum_{i=1}^2 (\int_{t_0}^t {\alpha}_{i} \cdot H(t) \cdot P(\tau) \cdot e^{\frac{\tau-t}{{\lambda}_i}} d\tau + D_{\rm{history},i} \cdot e^{\frac{t_{0}-t}{{\lambda}_i}}),
\end{equation}
where $t_{0}$ is the beginning time of the background light curve, $\alpha_i$ is the amplitude of generating the $i$th radioisotope, $\lambda_i$ is the corresponding decay constant, $D_{\rm{history},i}$ is the delayed background of the corresponding radioisotope leftover at $t_{0}$, $H(t)$ is the modulation coefficient of the satellite altitude:
\begin{equation}
  H(t) = \eta \cdot \frac{2 \cdot (h(t) - \bar{h})}{h_{\rm max} - h_{\rm min}},
\end{equation}
where $\eta$ is the altitude parameter with absolute value $<$1; $h_{\rm max}$, $h_{\rm min}$ and $\bar{h}$ are the maximum, minimum and mean values of satellite altitude $h(t)$, respectively.

In summary, the expression of the HE/CsI background parametric model is as follows:
\begin{equation}
  B(t)=\sum_{i=1}^2 (\int_{t_0}^t {\alpha}_{i} \cdot H(\tau) \cdot P(\tau) \cdot e^{\frac{\tau-t}{{\lambda}_i}} d\tau + D_{\rm{history},i} \cdot e^{\frac{t_{0}-t}{{\lambda}_i}}) + P(t) + a_{4}.
\end{equation}

During the burst of GRB 221009A and several hours around the burst, \insight/HE was making scan observations of the Galactic plane in LG Mode; there is no known bright celestial source in the scanned area, therefore this scan data is dominated by background, except the GRB episode. The earth eclipse time of \insight ~observing GRB 221009A is 1883$-$3864 s since $T_{0}$. Therefore, we choose LG Mode HE/CsI data in -4220$-$0 s and 2000$-$5780 s since $T_{0}$ (8 ks in total) as background to train the parameters of the background model. The counts of the 18 \insight/HE main detectors are added together to generate the light curves with time bin of 10 s and energy band width of 50 keV. Due to the large number of parameters in the model and to avoid local optimal solutions, we use Genetic Algorithm (GA) to search the values of the parameters, and the errors of the parameters are calculated by MCMC method. The error of background model can be calculated as follows:
\begin{equation}
  {\sigma^{2}_{\rm{bkg}}} = {\sigma^{2}_{\rm{ACD}}} + {\sigma^{2}_{\rm{para}}},
\end{equation}
where $\sigma_{\rm{bkg}}$ is the total error of background model, $\sigma_{\rm{ACD}}$ is the model error caused by the statistical error of ACD count rate, $\sigma_{\rm{para}}$ the error caused by the GA searching error of parameters. After GA training and background error calculating, we use the parametric background model to generate the HE/CsI net count rates in range of 400$-$2950 keV and 0$-$2000 s since $T_{0}$. The error of HE/CsI net count rate is:
\begin{equation}
  {\sigma^{2}_{\rm{net}}} = {\sigma^{2}_{\rm{bkg}}} + {\sigma^{2}_{\rm{CsI}}},
\end{equation}
where $\sigma_{\rm{net}}$ is the error of CsI net count rate, $\sigma_{\rm{CsI}}$ is the statistical error of CsI total count rate.

\begin{figure}
  \centering
  \includegraphics[width=15 cm]{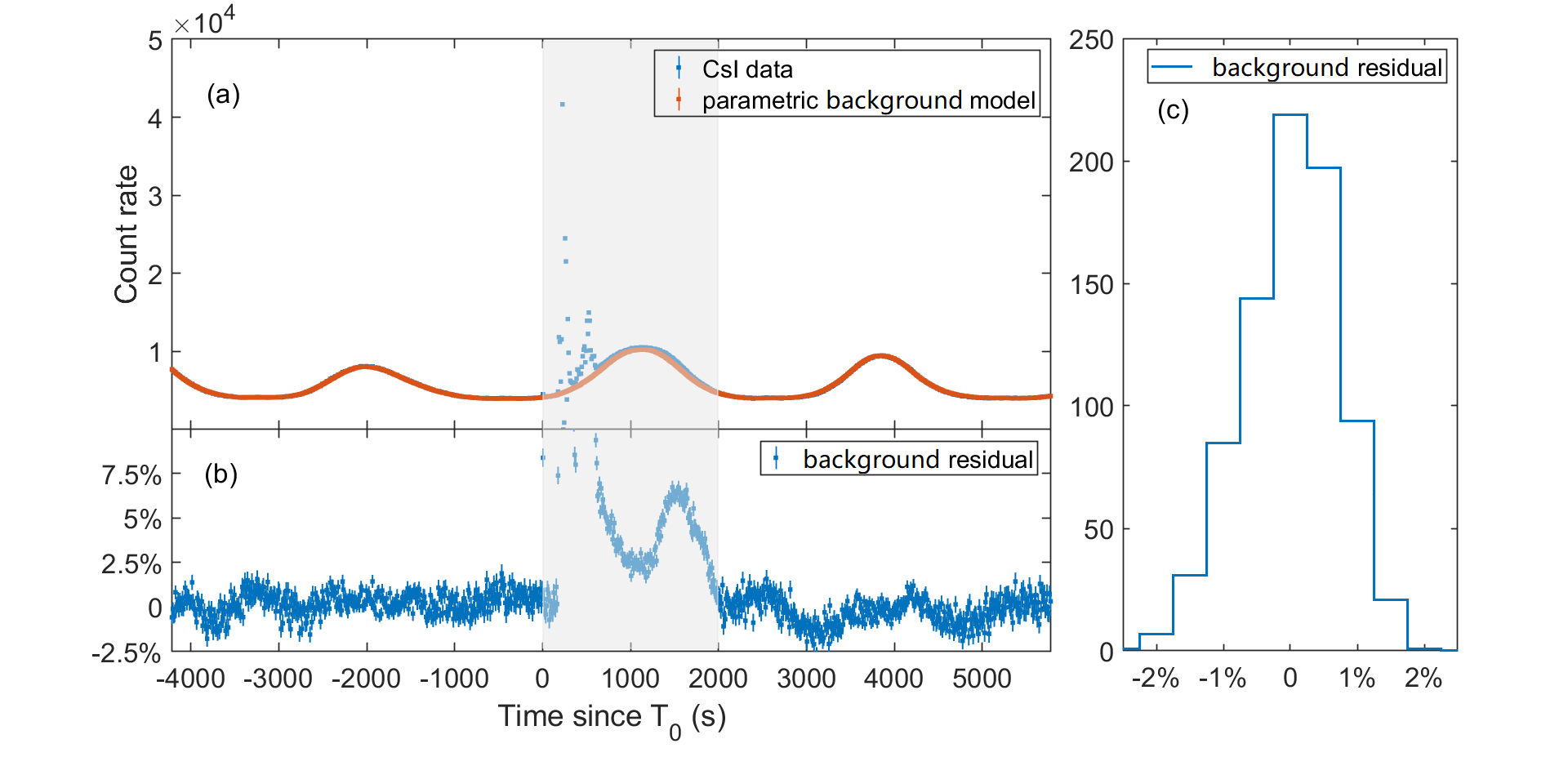}
  \caption{\insight ~HE/CsI background model as a function of time. Panel (a) shows light curves of HE/CsI count rate in 400$-$2950 keV and the corresponding HE/CsI background light curve; panel (b) is the relative residual of HE/CsI background model; panel (c) is the histogram of the relative residual of the 8-ks length training data. Data in the white area are employed in background model fitting and the gray area corresponds to the time range of the background to be estimated.}
  \label{figure:CsIbkg-time}
\end{figure}

\begin{figure}
  \centering
  \includegraphics[width=15 cm]{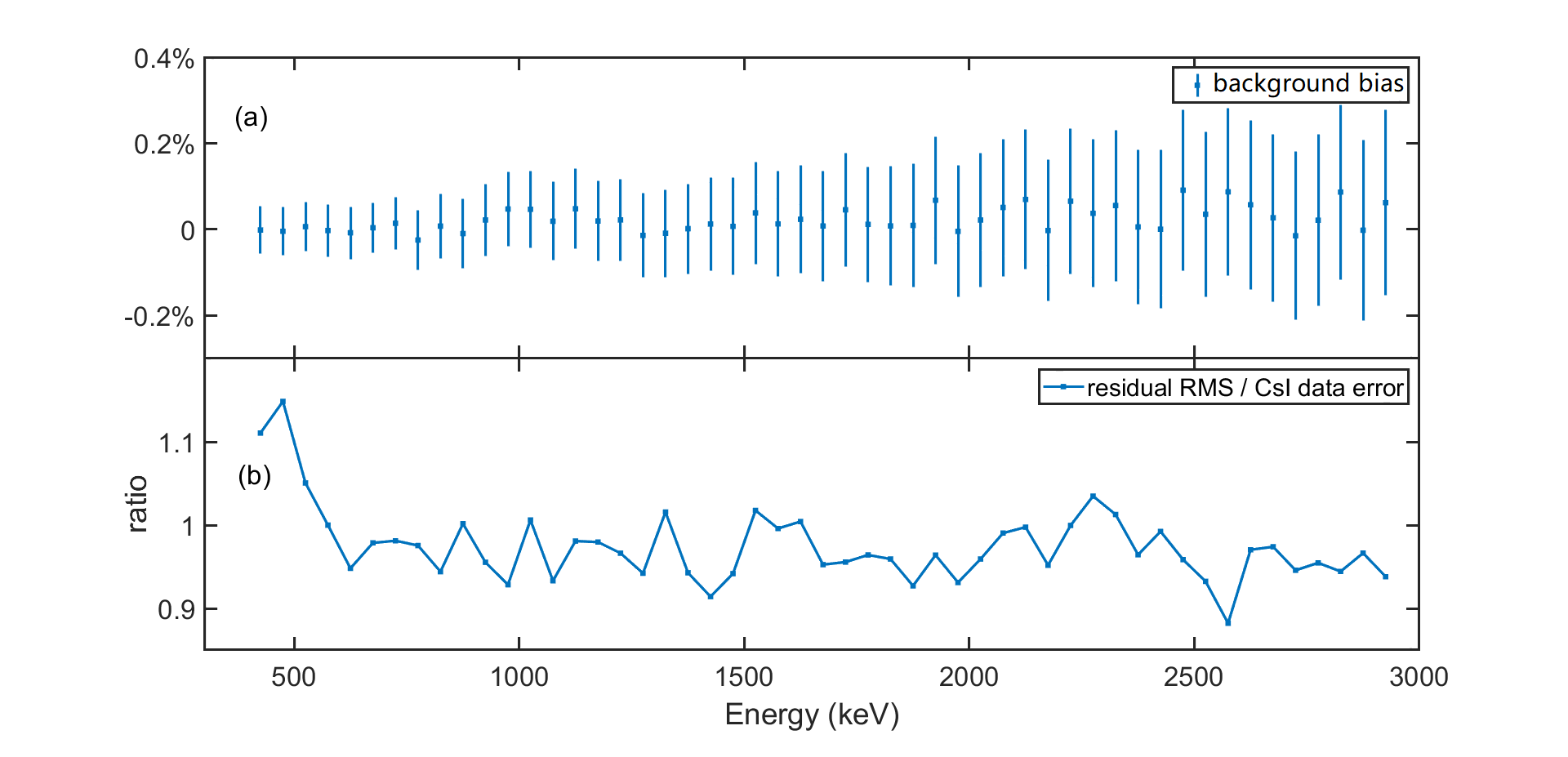}
  \caption{\insight ~HE/CsI background model as a function of energy of the 8-ks training data. Panel (a) shows the bias of the background model in different energy bands; panel (b) is the ratio between $\sigma_{\rm{CsI}}$ and the residual RMS of the background model alone energy.}
  \label{figure:CsIbkg-energy}
\end{figure}

The reliability of the HE/CsI background model is tested as functions of both time and energy. Figure \ref{figure:CsIbkg-time} shows that the background light curve is well fitted in the 8-ks length training range, and the background model has no obvious bias along time. The net count rate in 600$-$900 s since $T_{0}$ is obviously larger than the background residuals of the training data, indicating that the early afterglow of GRB 221009A is detected by HE/CsI data. Panel (a) of Figure \ref{figure:CsIbkg-energy} proves that the background model reproduces the spectrum of HE/CsI background without bias, and panel (b) suggests that the systematic error of the background model is slightly larger at energy $<$550 keV but can be neglected at energy $>$550 keV. Therefore, the HE/CsI net counts $<$550 keV generated by the parametric background model has larger error.

\begin{figure}
  \centering
  \includegraphics[width=15 cm]{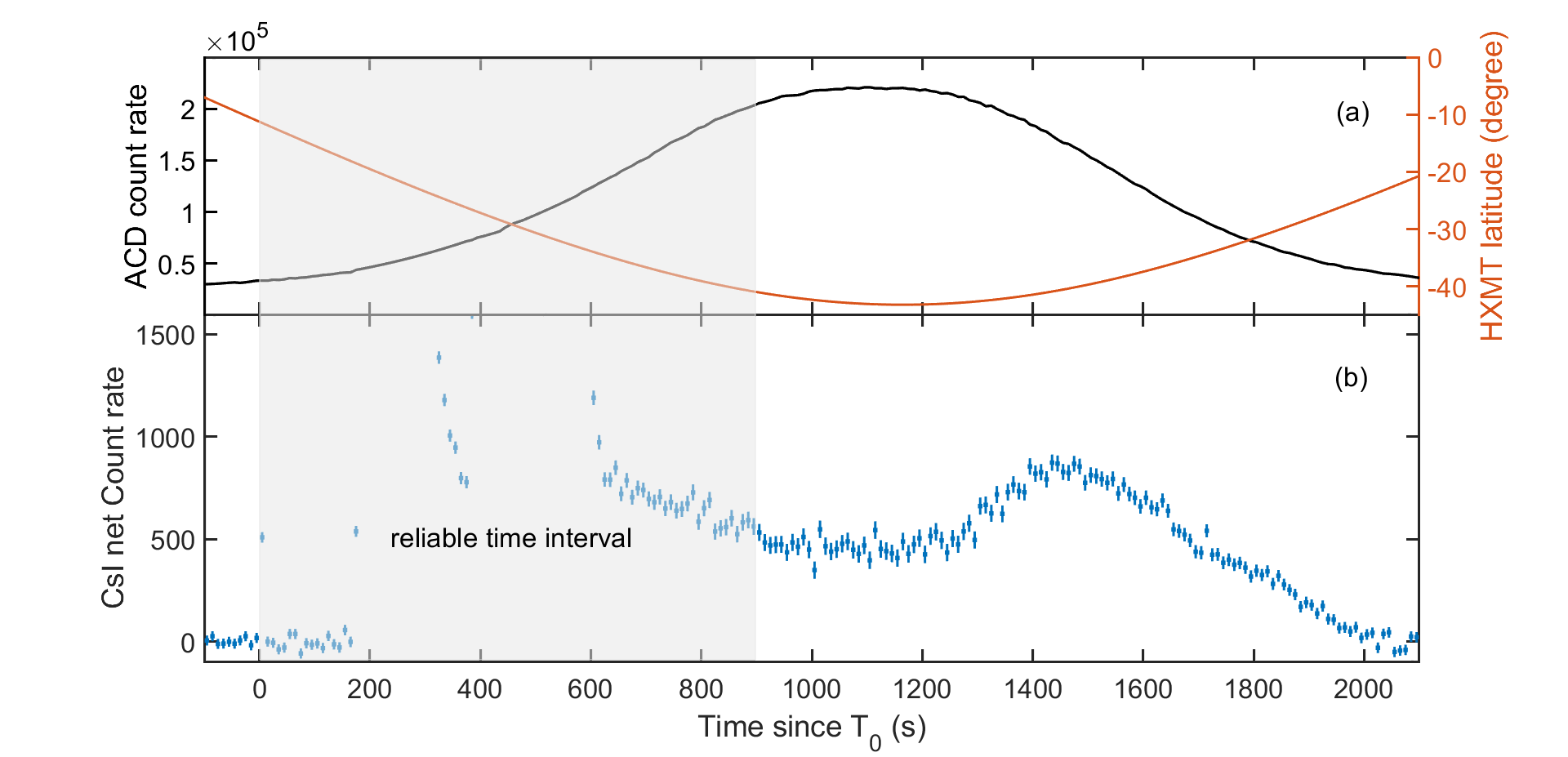}
  \caption{Panel (b) is the \insight ~HE/CsI net light curve in 400$-$2950 keV generated by the parametric background model in time range of 0$-$2000 s since $T_{0}$, while panel (a) shows the \insight~ latitude and ACDs count rate in the corresponding time range. We consider that the only net count rate of HE/CsI in the gray area (0$-$900 s since $T_{0}$) is reliable.}
  \label{figure:CsInetLC}
\end{figure}

The net light curve of 400$-$2950 keV between 0$-$2000 s since $T_{0}$ is drawn in Figure \ref{figure:CsInetLC}. The net light curve is around 0 between the precursor and the ME, further verifying the reliability of the background model. However, there is an abnormal peak in the net light curve after 1000 s since $T_{0}$. This is because the \insight~satellite suffers some particle events in high latitude region. These particle events occur randomly in the high latitude radiation zone, but are not well detected by ACDs. Therefore, the prompt backgrounds and delayed backgrounds of the particle events are not simulated by the parametric model, leading to the underestimation of the background when and after the satellite passing through the high latitude area. The orbit of \insight, shown in Figure \ref{fig:SatelliteTrack}, is close to the south latitude radiation zone. Therefore, to be conservative, we consider that the net counts $>$900 s since $T_{0}$ generated by the background model is not reliable enough, so that the afterglow accurately detected by \insight ~HE/CsI is $<$900 s since $T_{0}$.

\subsubsection*{GECAM-C background}

GECAM-C is onboard the Space advanced technology demonstration satellite (SATech-01) satellite, which is a sun-synchronous orbit satellite at an altitude of 500 km and an inclination of 97.4°. This orbit is not often used for astronomical satellites because of the complex space particle environment. The conventional method for background determination with wide-field gamma-ray monitor such as GECAM is to time-interpolate intervals before and after the source as a polynomial. However, for the long lived emission and high latitude area, it is difficult to estimate background of GRB 221009A by this method. A more feasible and reliable way is to use the revisited orbit which has approximately the same geographical footprint as this very orbit for background estimation. This method has been successfully applied in CGRO/BATSE \cite{2002ApJ...567.1028C} and \textit{Fermi}/GBM \cite{2012arXiv12105369}. The track of subsatellite point during the period of GRB 221009A is shown in Figure \ref{fig:SatelliteTrack}, while the two revisited orbits are also plotted, with one happened five days ago ($T_0$-431780s, plotted as yellow) and the other happened five days later ($T_0$+431750s, plotted as blue). Figure \ref{fig:SatelliteTrack} also indicates that the background of GECAM-C is very stable except during the part of the high latitude region, where GECAM-C fortunately went out when the burst occurred.

In order to estimate the background using the revisited orbit, the direction of the GRDs needs to be the same. The primary observation modes of SATech-01 are inertial observation of the sun and inertial observation of the lobster eye X-ray telescope (another payload on satellite). The direction of the satellite remains the same in both modes, except for mode switching, which causes the satellite to slew. The satellite had just completed the observation mode switching to inertial observation of the sun at the trigger time of \insight, the beginning of GRB 221009A, then maintained its inertial orientation until it was blocked by the Earth (Figure \ref{fig:IncidentAngle}). Thus the direction of the GRDs during the revisited orbit remained the same as that during the burst period.

\begin{figure}[htbp]
   \centering  
   \subfigure[GRD01 High Gain]{
   \label{fig:RawLC-GRDHG}
   \includegraphics[width=0.48\textwidth]{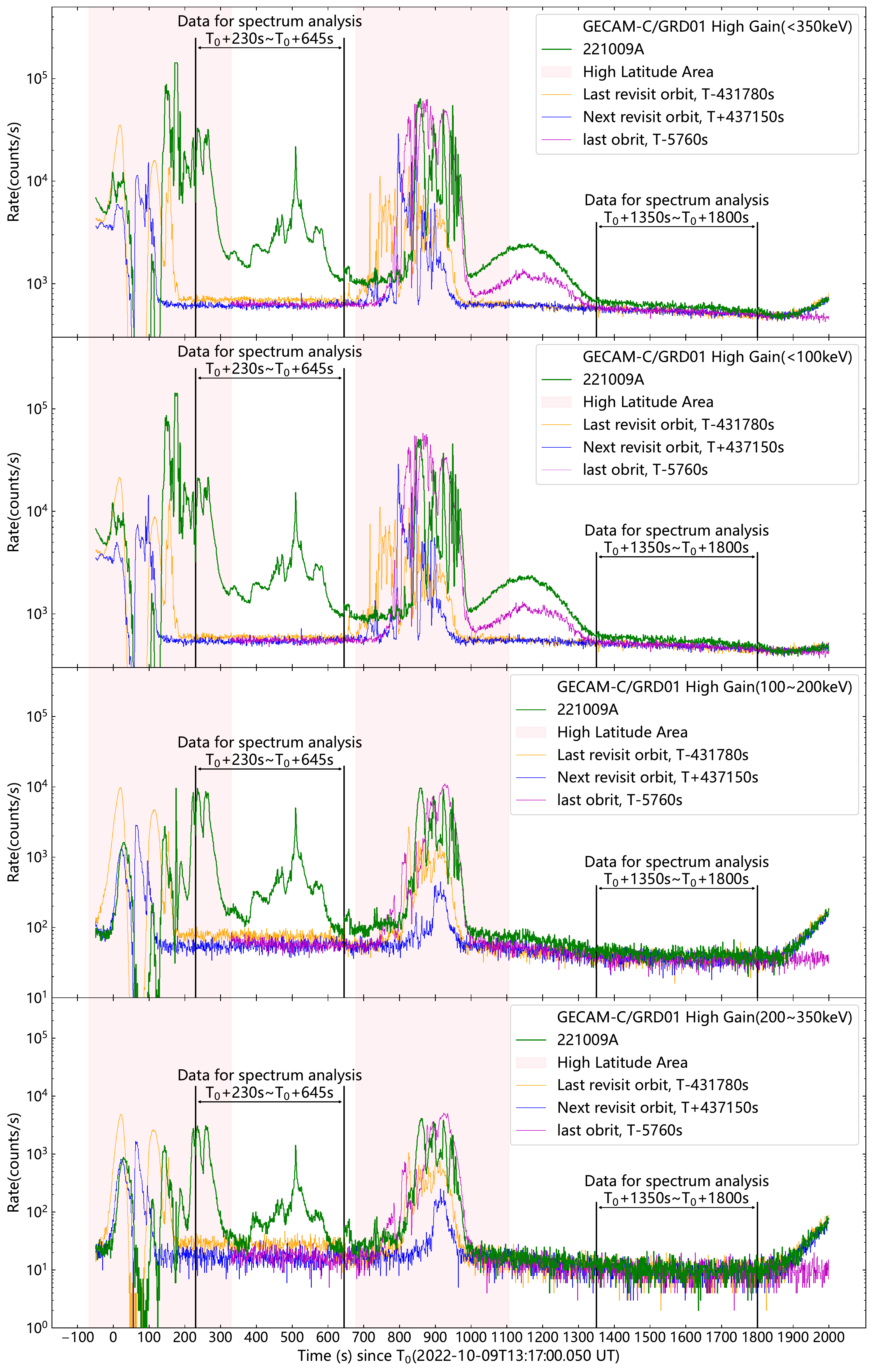}}
   \subfigure[GRD01 Low Gain]{
   \label{fig:RawLC-GRDLG}
  \includegraphics[width=0.48\textwidth]{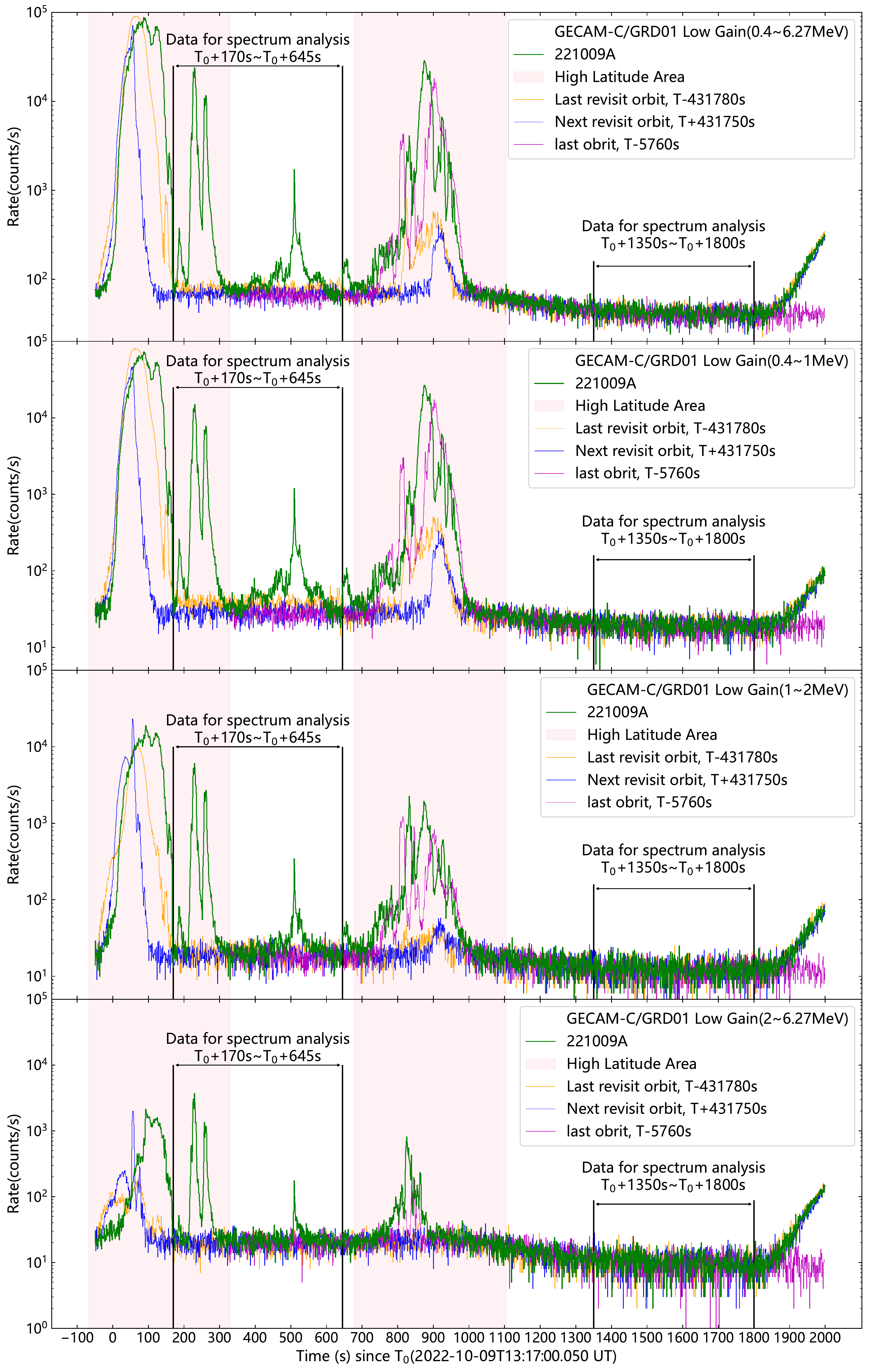}} 
   \caption{Raw light curves and background of GECAM-C. Panel (a) shows the GRD01 High Gain light curves. Panel (b) shows the GRD01 Low Gain light curves. The magenta areas are the high latitude areas. The green lines are the light curves of GRB 221009A. And other color lines are used to estimate backgroud.}
   \label{fig:RawLC}
\end{figure}

The background, that is, the spatial environment, is generally determined by geographical location of the satellite, but it will fluctuate with time. Therefore, when using the revisited orbit for background estimation, the fluctuations of space environment also need to be considered; hence the last orbit before the burst (plotted as purple in figure \ref{fig:RawLC}) is selected for comparison. 

\begin{figure}[htbp]
    \centering  
    \subfigure[$T_0$+600s$\sim$$T_0$+640s]{
    \label{fig:bkgspec_600}
    \includegraphics[width=0.48\textwidth]{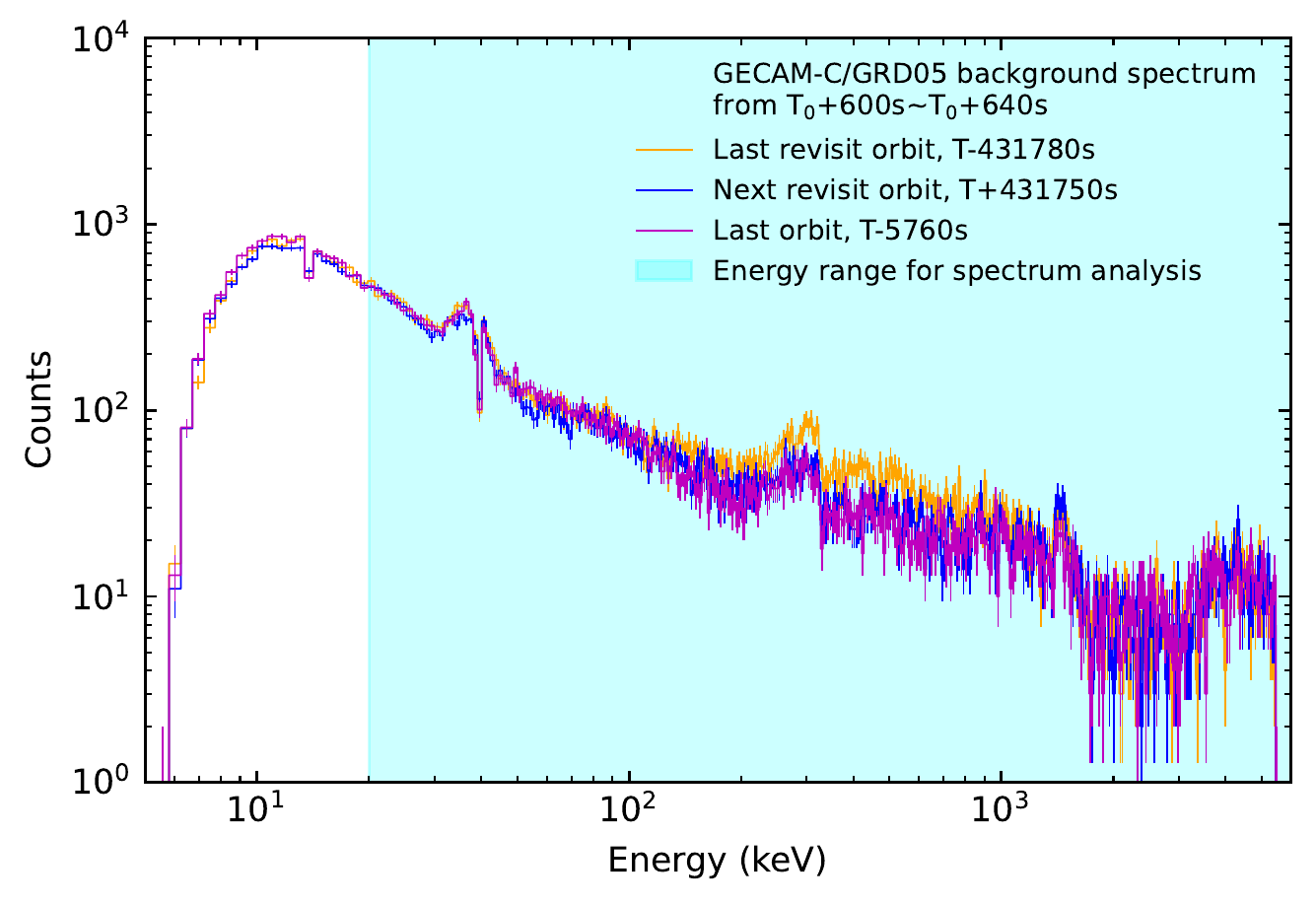}}
    \subfigure[$T_0$+1350s$\sim$$T_0$+1800s]{
    \label{fig:bkgspec_1300}
    \includegraphics[width=0.48\textwidth]{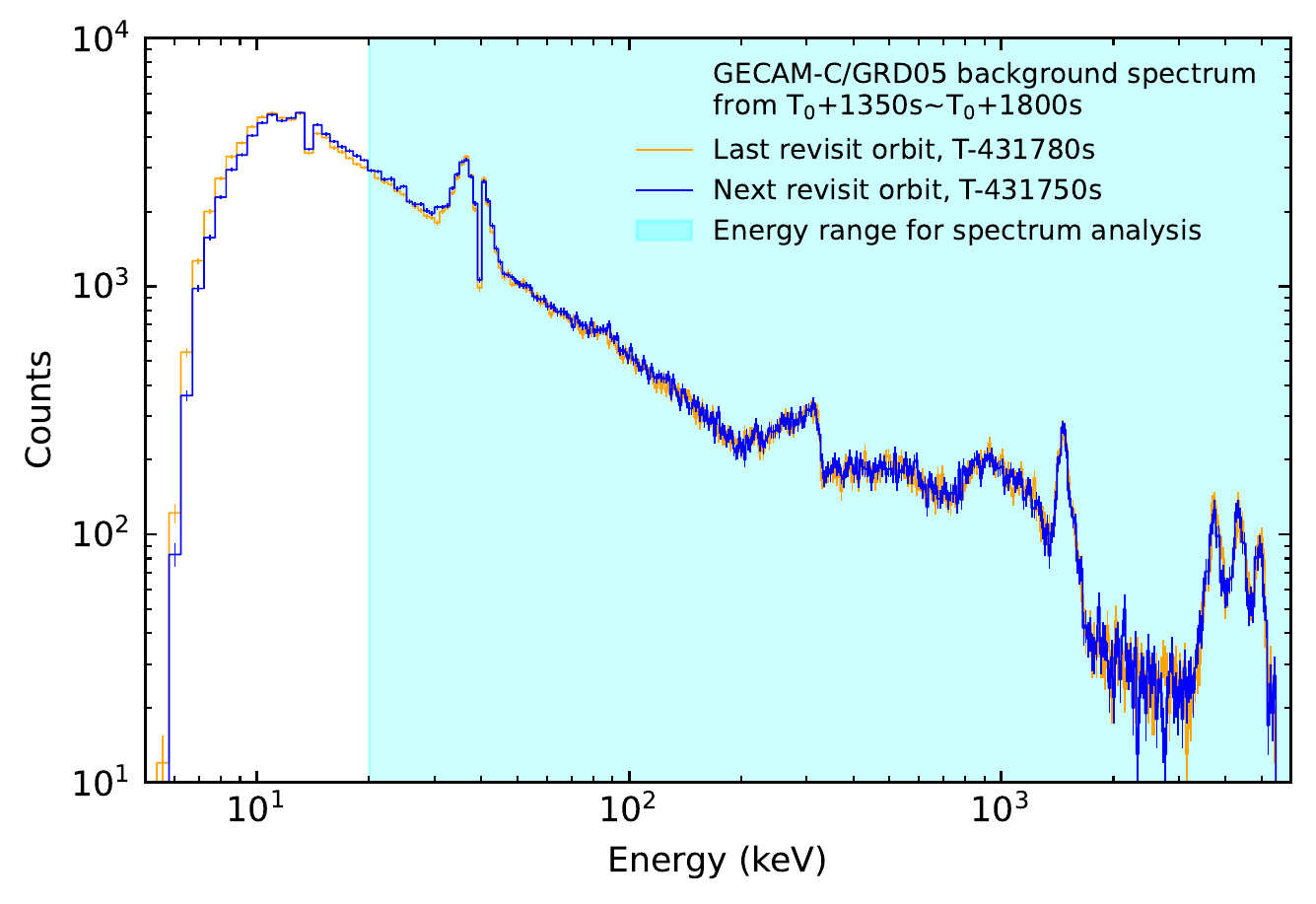}} 
    \caption{Spectrum of background of GECAM-C. Panel (a) shows the background spectrum for the time interval from $T_0$+600\,s to $T_0$+640\,s. Panel (b) shows the background spectrum for the time interval from $T_0$+1350\,s to $T_0$+1800\,s.}
    \label{fig:bkgspec}
\end{figure}

All the original light curves can be seen in Figure \ref{fig:RawLC}. The trend of the light curves of the two revisited orbits is consistent. GRD01 is the detector that kept working and fully recorded GRB 221009A during main peak. The average HG count rate difference between the two revisited orbits is about 80 counts/s (10 keV $\sim$ 350 keV) and the average LG count rate difference is about 10 counts/s (400 keV $\sim$ 6 MeV), which are three orders of magnitude weaker than the burst signal. During the afterglow, the average HG count rate difference between the two revisited orbits is about 3 counts/s (20 keV $\sim$ 140 keV), and the average LG count rate difference is about 0.17 counts/s (400 keV $\sim$ 6 MeV). GRD05, which has the smallest zenith to GRB 221009A, has observed more details of the afterglow; the counting spectra of both two revisited orbits are plotted in Figure \ref{fig:bkgspec}. In the energy band used for spectral analysis, counting spectra are consistent within the error range. The count rate difference between the two revisited orbits is about 24 counts/s (20 keV $\sim$ 140 keV) from $T_0$+600s to $T_0$+645s, and about 7 counts/s (20 keV $\sim$ 140 keV) from $T_0$+1350s to $T_0$+1800s.
Apart from that, both of the light curves of revisited orbits are very smooth in the time period for data analysis, which is not in conflict with the conclusion from Figure \ref{fig:SatelliteTrack}. There is a bump on the light curve of <100 keV when leaving the high latitude region at $T_0$+1100s and the last orbit, however, this structure does not appear on the light curve of the revisited orbit, which implies that a low energy space particle event occurred in this region on October 9, and the data of the last orbit can reflect the space environment at the time of the burst to some extent. Therefore, the revisited orbit five days later, whose background level is consistent with the last orbit before the burst, is considered to be a more suitable estimation of the space environment at the time of the burst; this is also verified by the Earth occultation analysis (see below). Hence the time of afterglow analysis is selected from $T_0$+1350s to $T_0$+1800s to minimize the effect of space particle event on the correctness of the results. 

\subsubsection*{Spectral Analysis}
The spectral model used in our analysis is Band (Eq.\ref{equ:band_Model}), cut-off power-law (Eq. \ref{equ:Cutoffpl_Model}) and power-law (Eq. \ref{equ:pl_Model}).
\begin{equation}
    N_{\rm Band}(E)= \begin{cases} 
    A \bigg (\frac{E}{E_{\rm piv}}\bigg )^{\alpha} {\rm exp}\bigg(-\frac{E}{E_{\rm c}}\bigg ), &\quad (\alpha -\beta)E_{\rm c} \geq E, \\ 
    A \bigg [\frac{(\alpha-\beta) E_{\rm c}} {E_{\rm piv}}\bigg ]^{\alpha-\beta} {\rm exp}(\beta-\alpha)\bigg (\frac{E}{E_{\rm piv}}\bigg )^{\beta}, &\quad (\alpha -\beta)E_{\rm c} \leq E, \\
    E_{\rm piv} = 100 \enspace {\rm keV},
    \end{cases},
    \label{equ:band_Model}
\end{equation}
where $A$ is the normalization amplitude constant ($\rm photons \cdot cm^{-2} \cdot s^{-1} \cdot keV^{-1}$), $\alpha$ and $\beta$ are the \textit{low-energy} and \textit{high-energy} power law spectral indices, $E_{\rm c}$ is the characteristic energy in keV, $E_{\rm piv}$ is the pivot energy in keV and nominally fixed to 100 keV.
\begin{equation}   
N_{\rm Cut-off \thinspace pl}(E)=AE^{-\alpha} {\rm exp}(-\frac{E}{E_{\rm c}}),
\label{equ:Cutoffpl_Model}
\end{equation}
where $A$ is the normalization amplitude constant ($\rm photons \cdot cm^{-2} \cdot s^{-1} \cdot keV^{-1}$), $\alpha$ is the power law photon index, and $E_{\rm c}$ is the characteristic cutoff energy in keV. 
\begin{equation}   
N_{\rm power-law}(E)=AE^{-\alpha},
\label{equ:pl_Model}
\end{equation}
where $A$ is the normalization amplitude constant ($\rm photons \cdot cm^{-2} \cdot s^{-1} \cdot keV^{-1}$), $\alpha$ is the dimensionless photon index of power-law.

\begin{figure}
\includegraphics[width=\textwidth]{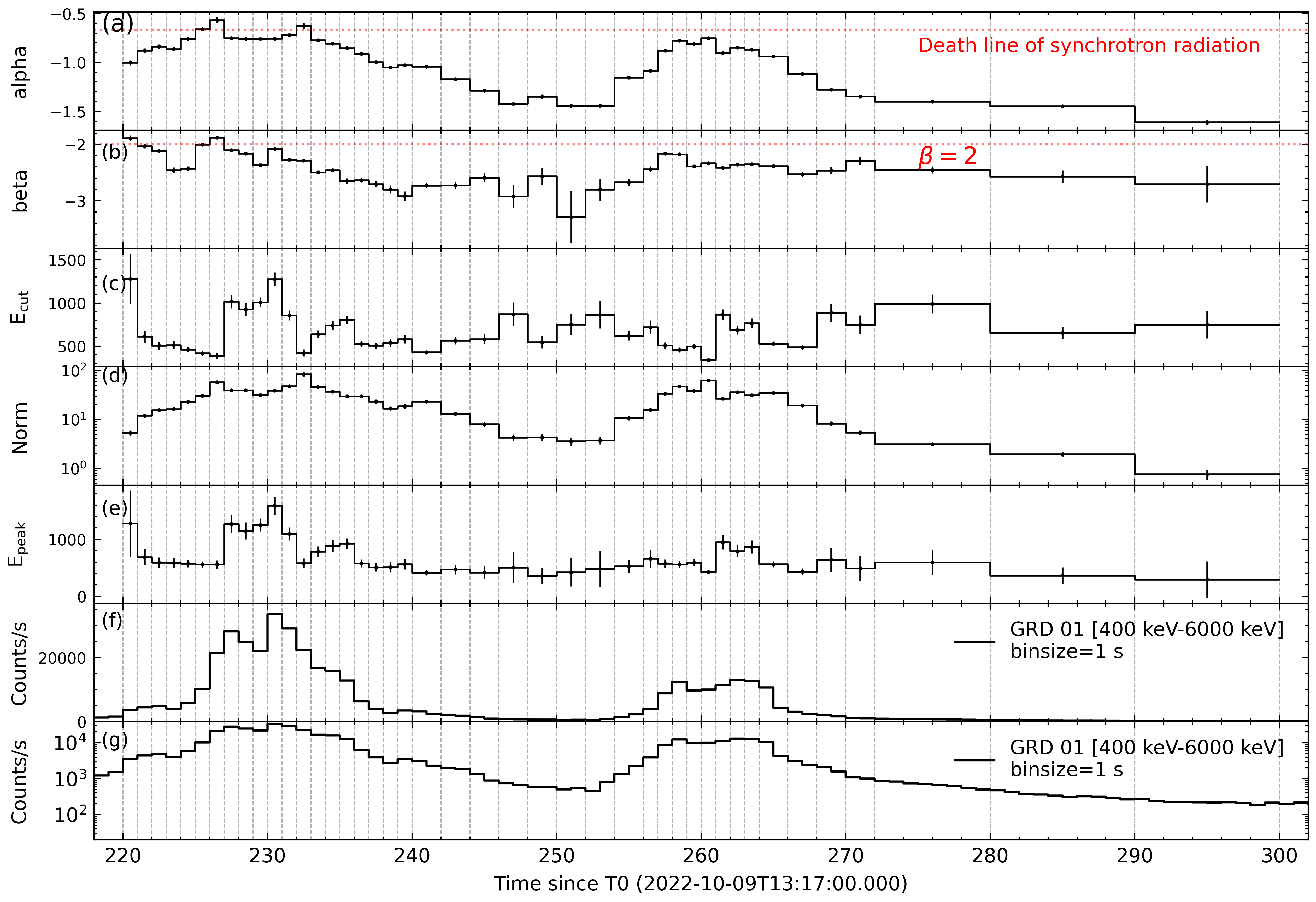}
\caption {Spectral evolution during the ME of GRB 221009A. Panel (a): low-energy spectrum index $\rm \alpha$ for the band model; Panel (b): high-energy spectrum index $\rm \beta$ for the band model; Panel (c): characteristic energy in keV; Panel (d): the normalization amplitude constant ($\rm photons \cdot cm^{-2} \cdot s^{-1} \cdot keV^{-1}$); Panel (e): peak energy in $\rm \nu \mathcal {F}_{\nu}$, and $\rm E_{peak} = (2+\alpha)\cdot E_{cut}$; Panel (f): Low-gain light curve for GECAM-C/GRD01, and the y-axis is a linear coordinate; Panel (g): Low-gain light curve for GECAM-C/GRD01, and the y-axis is a logarithmic coordinate.}
\label{fig:SpecEvo}
\end{figure}

For GRB 221009A shortly before the ME until near the end of the flare (from T = 2022-10-09T13:17:00.000 + 180 s to T + 600 s), because of the high latitude effect, we only have {\it Bspec} and {\it Btime} data, so the time starting point of spectral analysis is slightly different from the trigger time $\rm T_{0}$.
In this time range, we use the {\it Bspec} data of the GECAM-C/GRD01 detector for spectral fitting. It is important to note that for the time periods where HG dead-time recording is problematic, we free the constant factor for HG to correct for this effect. We have performed a detailed time-resolved spectral fitting and here we show a plot of the evolution of the spectral parameters over time in the ME phase in Figure \ref{fig:SpecEvo}.

\begin{figure}
\centering
\includegraphics[width=0.8\textwidth]{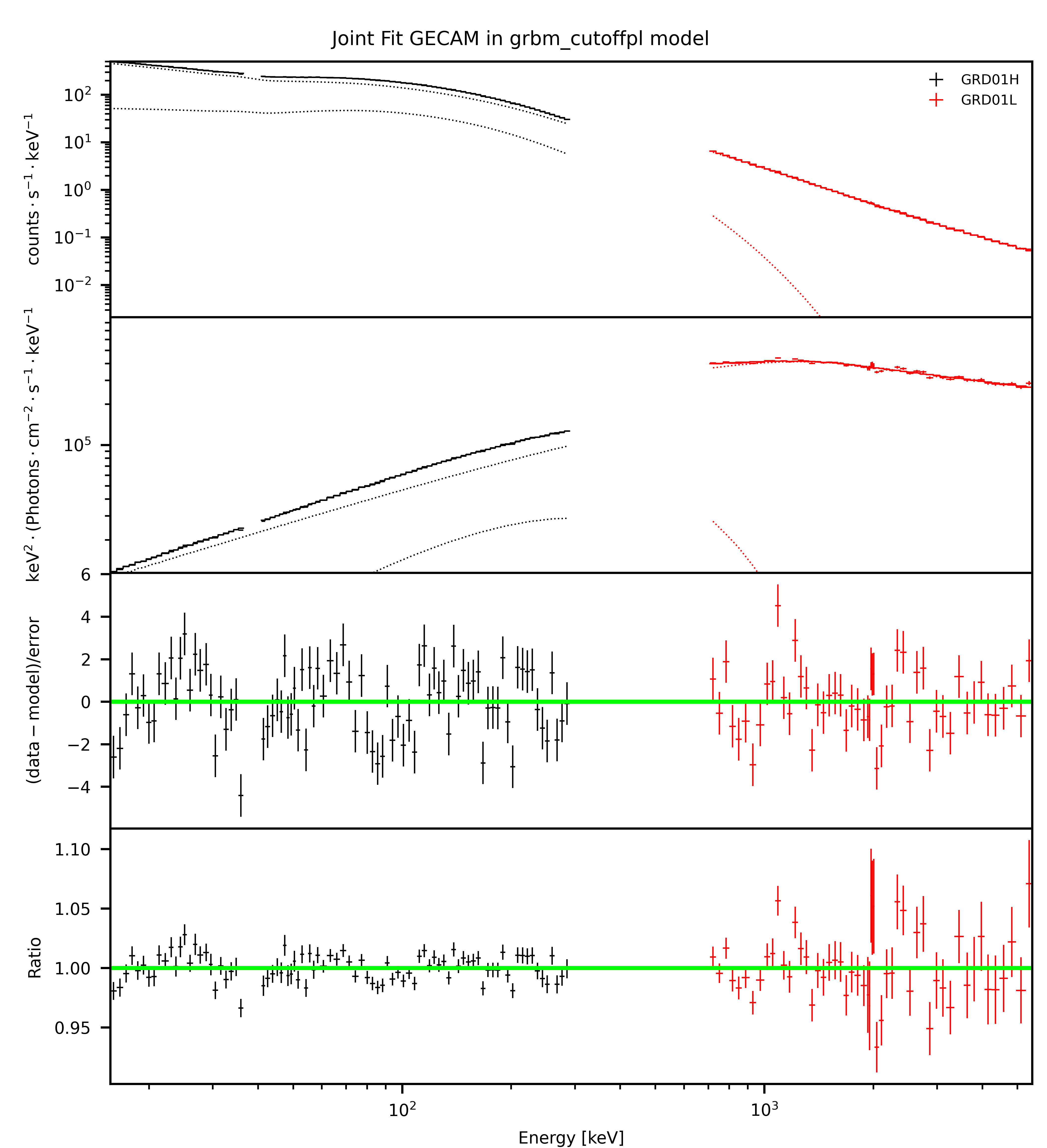}
\caption {Spectral fitting of the ME. The time interval is $T_0$+(220,300) s, a cut-off power-law component is added to eliminate the evolution of the spectrum. And the black data points are HG, the red data points are LG.}
\label{fig:specofmb2}
\end{figure}

In addition, we divide the ME into two peaks, which are presented in the upper panel of Figure \ref{fig:AmatiRelation}. We have done the time-integrated spectral fitting of the whole ME as well as these two peaks, and they all fit very well. The fitting results of Peak-1 are shown in Figure \ref{fig:specofmb2} with reasonable fitting residuals; we also add a cut-off power law component to accommodate possible fast evolution of spectral parameters. After the spectral fitting we calculate the isotropic energy of GRB 221009A for the full burst, ME, Peak-1, Peak-2, and Flare; the results are summarized in Table \ref{Eiso result}, where the data points are plotted on the Amati-Relationship diagram.

In addition, there is a high count rate effect for the brightest time segments of ME phase. Thankfully, there is only some slight effect on the HG during the brightest period (listed in the Table \ref{high_count_rate_effects}). Most importantly, it only affects the energy range below 100 keV. Therefore, we perform spectrum fitting for the case of ignoring below 100 keV and keeping it, respectively, and the results show that the high energy spectrum index and energy flux are almost unchanged. For the sum of the $E_{\rm iso}$ of these bins, the result calculated by ignoring the values below 100 keV changed by less than 4 percent compared with the original result. This means that the high count rate effect can be ignored in the overall analysis.

\begin{table}
\centering
 \caption{{$E_{\rm iso} $ of GRB 221009A Full burst, ME, Peak-1, Peak-2 and Flare}}
% \begin{supertabular*}{}{@{}@{\extracolsep{\fill}}cccc@{}}
 \resizebox{\textwidth}{!}{\begin{tabular}{cccc}
\hline \hline
Time Interval & $E_{\rm iso} $ (erg) & $E_{\rm iso}$ (erg) with 10\% sys. err & $E_{\rm peak}$ (keV)\\
\hline
Full Burst & $(1.49\pm 0.01) \times 10^{55}$& $(1.5\pm 0.2) \times 10^{55} $ & ${1247.4} \pm {91.2}$\\
ME &$({1.33} \pm {0.02}) \times 10^{55} $& $({1.3} \pm {0.1}) \times 10^{55} $ & ${1247.4} \pm {91.2}$\\
ME Peak-1 & $({9.45} \pm {0.16}) \times 10^{54} $  & $({9.4} \pm {1.1}) \times 10^{54} $& ${1266.0} \pm {98.4} $ \\
ME Peak-2 & $({3.86} \pm {0.09}) \times 10^{54} $ & $({3.9} \pm {0.5}) \times 10^{54} $ & ${1152.2} \pm {170.0} $ \\
Flare & $({4.00} \pm {0.20}) \times 10^{53} $& $({4.0} \pm {0.6}) \times 10^{53} $ &${357.4} \pm {18.0} $\\
\hline \hline
% \end{supertabular*}
\label{Eiso result}
\end{tabular}}
\footnotesize{\leftline{Note: $E_{\rm iso}$ is calculated in energy range 1-10000/(1+$z$) keV;}}\\
\end{table}

\begin{table}
\centering
% \normalem 
\caption{{spectral fitting for time periods with high count rate effects}}
 \resizebox{\textwidth}{!}{\begin{tabular}{ccccccccc}
%  \begin{supertabular*}{\hsize}{@{}@{\extracolsep{\fill}}ccccccc@{}}
\hline \hline
  \multirow{2}*{Time range} & \multicolumn{3}{c}{High gain was reserved below 100keV} & \multicolumn{3}{c}{High gain is ignored below 100keV} &  \multicolumn{2}{c}{\multirow{2}*{Flux($\rm erg/cm^2/s$)}} \\ 
  \cline{2-7}
  
& $\alpha$ & $\beta$ & $E_{\rm cut}$ (keV)  & $\alpha$ & $\beta$ & $E_{\rm cut}$ (keV) & &\\
\hline
(227,228) s & $-0.75_{-0.02}^{+0.07}$ & $-2.10_{-0.03}^{+0.06}$ & $1015.66_{-331.79}^{+96.58}$ & $-0.91_{-0.06}^{+0.05}$ & $-2.16_{-0.04}^{+0.03}$ & $1400.87_{-131.99}^{+148.56}$  & $1.39_{-0.04}^{+0.18}  \times 10^{-2} $ & $1.39_{-0.03}^{+0.04}  \times 10^{-2} $  \\
(228,229) s & $-0.76_{-0.02}^{+0.02}$ & $-2.16_{-0.03}^{+0.03}$ & $924.80_{-84.07}^{+76.30}$ & $-0.70_{-0.06}^{+0.06}$ & $-2.18_{-0.03}^{+0.03}$ & $929.40_{-98.52}^{+88.02}$  & $1.21_{-0.03}^{+0.04}  \times 10^{-2} $ & $1.18_{-0.03}^{+0.04}  \times 10^{-2} $   \\
(229,230) s & $-0.76_{-0.01}^{+0.01}$ & $-2.37_{-0.04}^{+0.04}$ & $1009.24_{-68.94}^{+69.72}$ & $-0.91_{-0.05}^{+0.04}$ & $-2.44_{-0.06}^{+0.04}$ & $1285.16_{-79.91}^{+109.47}$  & $9.85_{-0.26}^{+0.33}  \times 10^{-3} $ & $9.96_{-0.25}^{+0.27}  \times 10^{-3} $   \\
(230,231) s & $-0.76_{-0.02}^{+0.02}$ & $-2.08_{-0.03}^{+0.03}$ & $1277.70_{-82.90}^{+85.47}$ & $-0.93_{-0.08}^{+0.06}$ & $-2.10_{-0.04}^{+0.03}$ & $1607.66_{-141.70}^{+231.04}$ & $1.72_{-0.03}^{+0.04}  \times 10^{-2} $ & $1.77_{-0.04}^{+0.06}  \times 10^{-2} $   \\
(231,232) s & $-0.72_{-0.02}^{+0.03}$ & $-2.28_{-0.03}^{+0.03}$ & $856.24_{-91.92}^{+56.53}$ & $-0.81_{-0.07}^{+0.06}$ & $-2.30_{-0.03}^{+0.03}$ & $986.38_{-81.17}^{+109.14}$  & $1.40_{-0.04}^{+0.09}  \times 10^{-2} $ & $1.40_{-0.04}^{+0.05}  \times 10^{-2} $  \\
(232,233) s & $-0.63_{-0.04}^{+0.02}$ & $-2.29_{-0.02}^{+0.02}$ & $422.90_{-28.98}^{+62.93}$ & $-0.87_{-0.05}^{+0.05}$ & $-2.37_{-0.03}^{+0.03}$ & $926.01_{-76.45}^{+84.34}$ & $1.27_{-0.09}^{+0.05}  \times 10^{-2} $ & $1.07_{-0.03}^{+0.04}  \times 10^{-2} $ \\
(233,234) s & $-0.77_{-0.02}^{+0.02}$ & $-2.50_{-0.03}^{+0.03}$ & $638.75_{-71.12}^{+64.06}$ & $-0.85_{-0.04}^{+0.04}$ & $-2.60_{-0.05}^{+0.05}$ & $867.49_{-73.40}^{+71.13}$  & $8.20_{-0.42}^{+0.59}  \times 10^{-3} $ & $7.54_{-0.25}^{+0.31}  \times 10^{-3} $  \\
\hline \hline
\label{high_count_rate_effects}
\end{tabular}}
\footnotesize{\leftline{Note: Flux is calculated in energy range 1-10000/(1+$z$) keV;}}\\
\end{table}

% Time-integrated spectral fits

\subsubsection*{GECAM-C measurement of the ME}
We have developed a simulation program to simulate the process of converting photon energy deposition into events data. This program creates pulse waveform samples and simulates the process of electronic judgment and selection of event signals. By inputting the count rate and energy spectrum information of the deposition events into the simulation, we can obtain the count rate and energy spectrum information after electronic processing. The resulting processed events data then undergoes data transmission, is downloaded to the data storage module, and is ultimately transmitted to the ground.
The parameters used in the simulation are based on actual measured values, and the specific electronic design is based on the GECAM electronic design paper\cite{https://doi.org/10.48550/arxiv.2112.04786}.

\begin{figure}
\centering  
\includegraphics[width=0.8\textwidth]{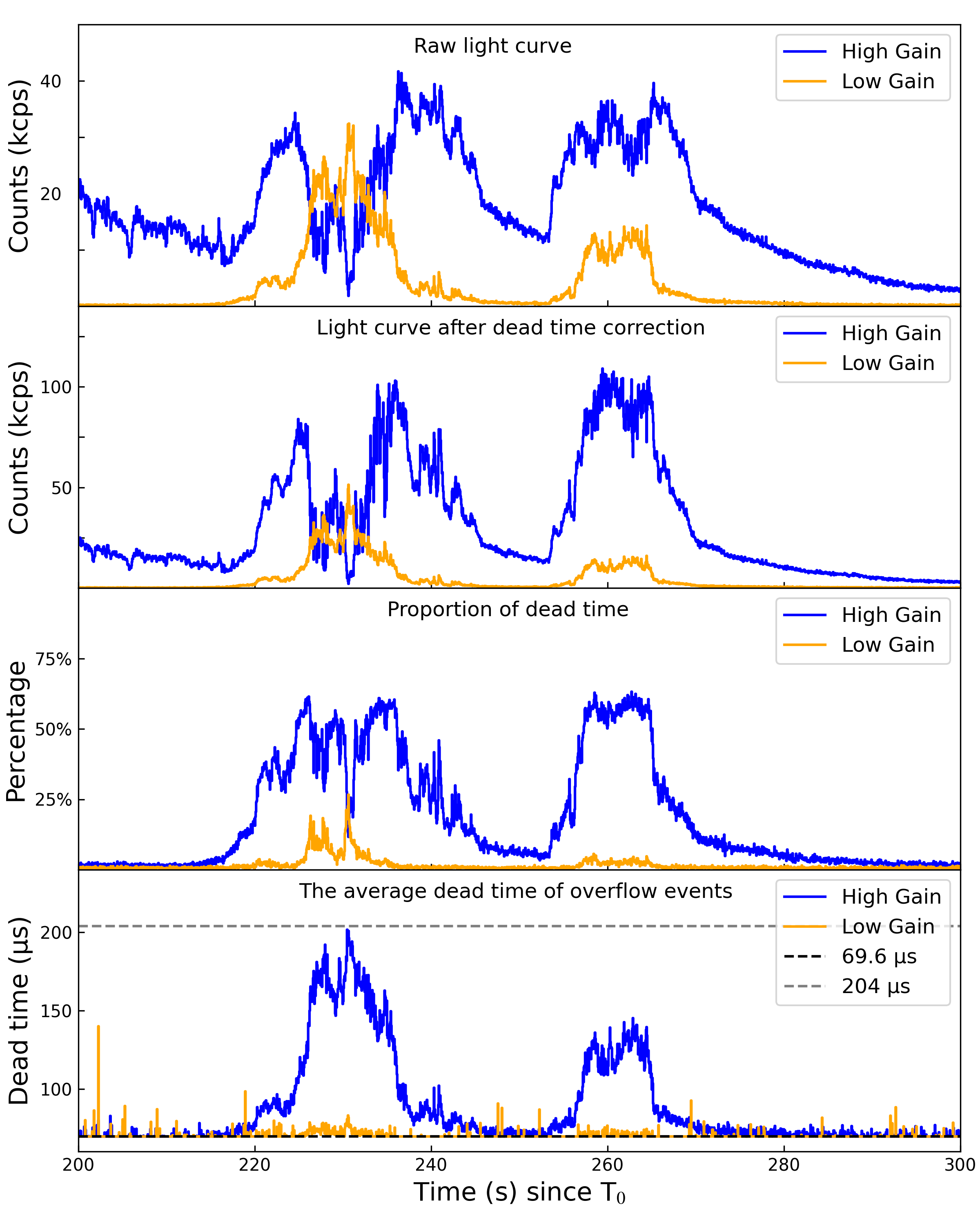}
\caption {Light Curve and dead time of GRD1.}
\label{fig:highRateCheck1}
\end{figure}

\begin{figure}
\centering  
\includegraphics[width=0.9\textwidth]{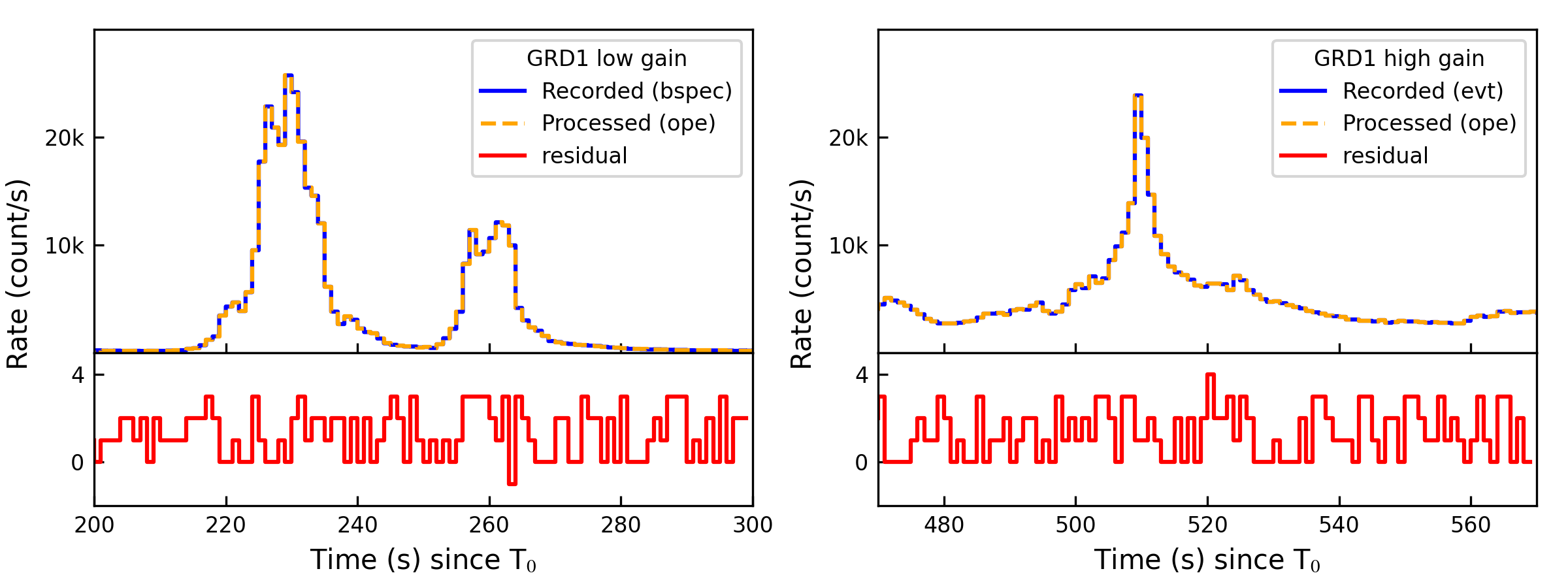}
\caption {Comparison between the processed with recorded count rates.}
\label{fig:highRateCheck2}
\end{figure}

\begin{figure}
\centering  
\includegraphics[width=0.9\textwidth]{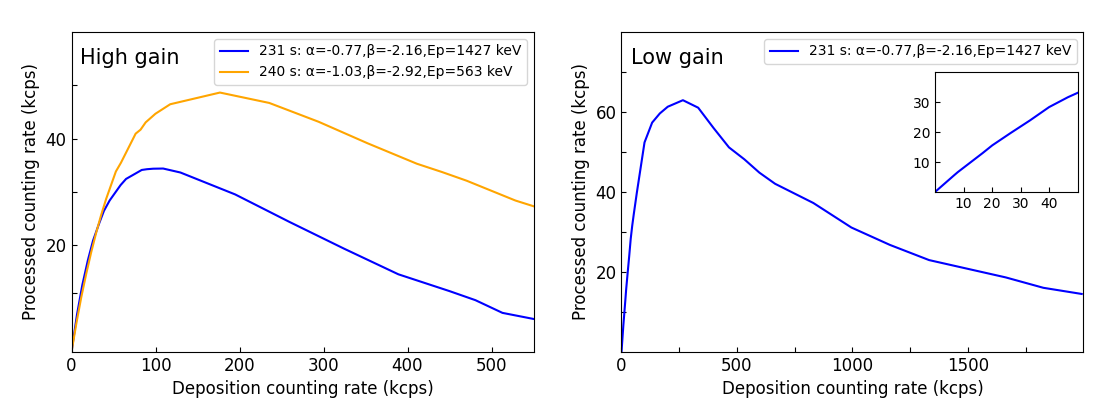}
\caption {The relationship between processing and deposition count rates obtained by simulation.}
\label{fig:highRateCheck4}
\end{figure}

For GECAM-C, the dead-time of normal events is 4 $\mu$s, while the dead-time of overflow events spans 69.6 $\mu$s to 204 $\mu$s, with the latter one is the maximum value (upper limit) that the data can record. 
When the incident count rate is high and the energy spectrum is hard, the observed count rate will decline.
When the count rate is very high, the dead-time of one event may be prolonged, and eventually exceeds the upper limit of the dead-time is recorded as 204 $\mu$s. 
At this time, the dead-time statistics are distorted and the observed count rate will also decrease (see Figure \ref{fig:highRateCheck4}). The average dead-time of overflow events can help us identify whether there is dead-time caused statistical distortion. During the ME, the dead-time of HG has increased significantly, indicating that there are a large number of events reaching the upper limit of dead-time, and the dead-time statistics are distorted.
However, the LG dead-time only slightly increased, indicating that the record dead-time of LG are still reliable (see Figure \ref{fig:highRateCheck1}). It is the same as the processing count rate in the engineering data, indicating that there is no saturation loss during the ME (see Figure \ref{fig:highRateCheck2}). The light curve of LG is thus reliable.

\begin{figure}
\centering  
\includegraphics[width=0.9\textwidth]{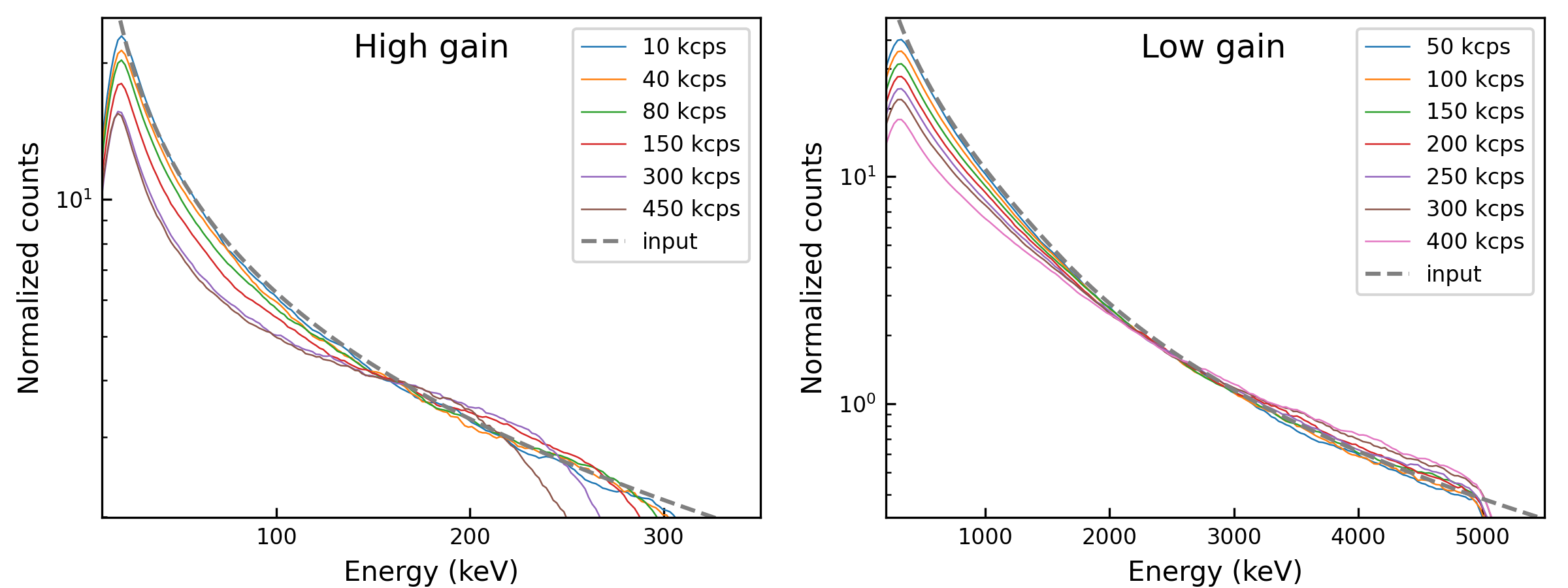}
\caption {Comparison between the original and the simulated GECAM-C energy spectra with different count rates.}
\label{fig:highRateCheck3}
\end{figure}

In the case of high count rate, energy spectrum distortion will occur, resulting in the difficulty of energy analysis; so it is necessary to analyze the energy spectrum distortion. For LG (0.4-6 MeV), the count rate is not very high (about 30 kcps). According to our simulation, it is necessary to reach an incident count rate of 100 kcps to produce significant energy spectrum distortion. GECAM-C adopts the dynamic baseline deduction mode, so the LG energy spectrum is reliable. However, for HG channel the rough estimation of the incident count rate can reach 300 kcps. With many high energy events, the energy spectrum distortion is relatively significant. From simulations we find that the spectral shape in the high energy band just has small distortion (see Figure \ref{fig:highRateCheck3}). This is because that when the accumulation occurs, unless the high energy event is accumulated with another high energy event, the energy change of this high energy event is very small compared with its energy. We thus believe that the energy spectral shape of high energy band data (>100 keV) of HG channel is reliable without significant pulse pileup effect; however, due to the incorrect record of dead time, we need to multiply a correction factor to get its ``true" flux.

\subsubsection*{Earth Occultation analysis for GECAM}
\begin{figure}
\centering
\includegraphics[width=0.8\textwidth]{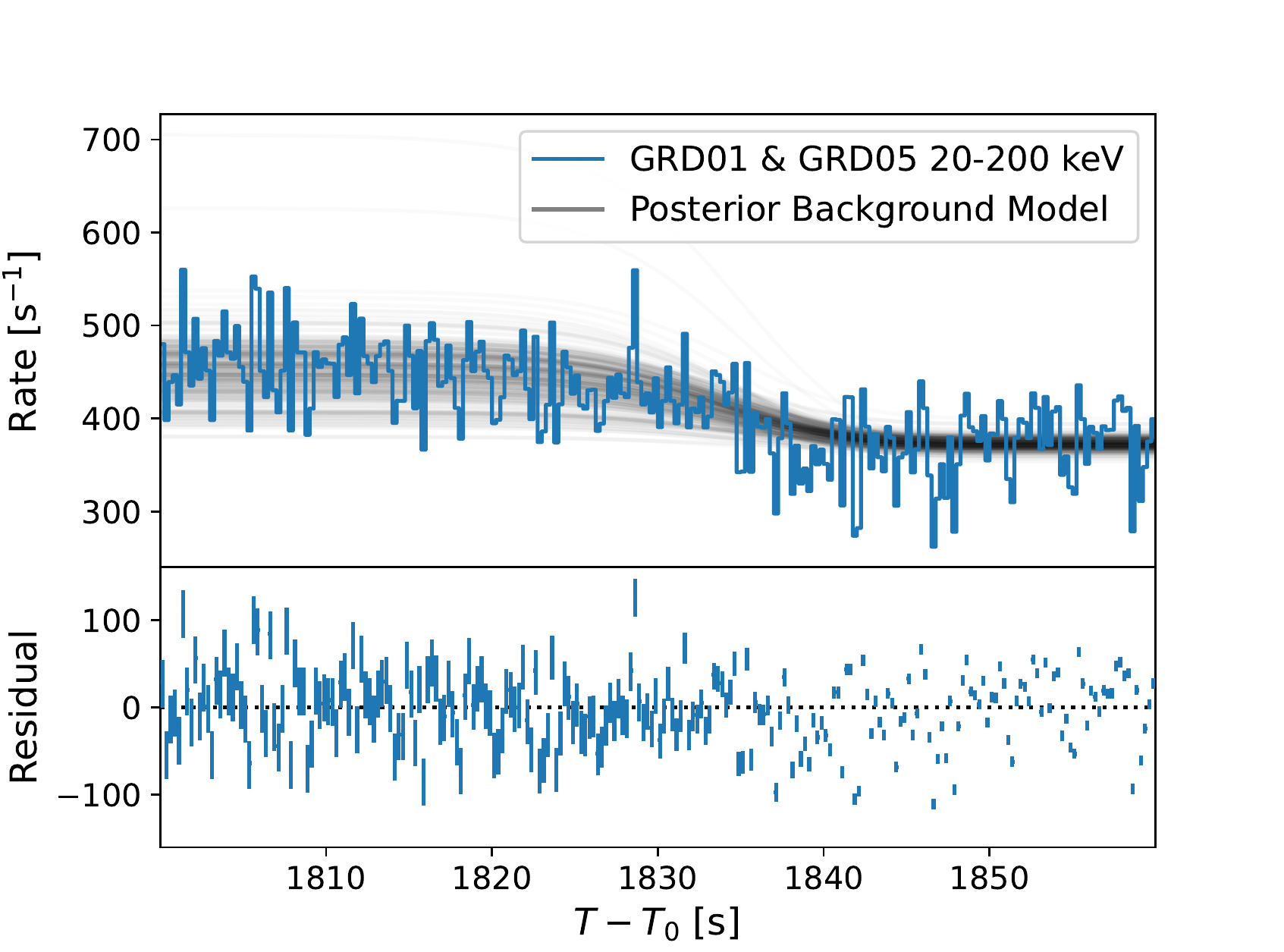}
\caption {Occultation light curve of GECAM-C/GRD01 \& GRD05. The gray lines are the light curves extracted from the posterior model. The lower end, mid, and upper end of the residuals are the difference of the observed light curve between 15.9\%, 50\%, and 84.1\% of the model values, respectively.}
\label{fig:OccultLC}
\end{figure}

For full-sky FoV gamma-ray monitor (e.g., GECAM), the data acquired after a source is occulted by the Earth or before the source is out of occultation can be used to estimate the background of the data when the source is unblocked. The occultation analysis are widely used to monitor the flux of X/gamma-ray source, and has been performed for data archive of many satellites \cite{Harmon2002OccultationTechnique,Harmon2004OccultationCatalog,Wilson-Hodge2012OccultationCatalog,Singhal2021OccultationCZTI}. Therefore, a relatively accurate spectral measurement of the occulted source can be obtained through Earth occultation analysis. Viewed from GECAM-B, GRB 221009A is unblocked by the Earth starting at $T_0$+1595\,s. However, the variation of particle background is complicated around this time, which needs to be further investigated, and hence no source estimation could be made from GECAM-B's data currently. For GECAM-C, GRB 221009A begins to be occulted by the Earth at $T_0$+1780\,s. The occultation light curve is shown in Figure \ref{fig:OccultLC}.

The afterglow signals are significant within the energy of about 20-200 keV for GECAM-C, and the spectral shape is simply a power-law; thus, the same energy range (except for 32-42 keV, where there exists the absorption edges) and spectral model will be used in the following occultation analysis for GECAM-C. The occultation data we analyze here are taken from GRD01 and GRD05, and the time range is from $T_0+1800$ s to $T_0+1865$ s. 
It is reasonable to assume that the spectral properties of the afterglow remain the same during the occultation process, and thus the source flux observed by GECAM-C can be modeled as
\begin{equation}
    s(A, \gamma, t, E_\mathrm{ch}) = \int {\rm d}E\ A E^\gamma\ {\rm e}^{-\tau(t, E)}\ R(E, E_\mathrm{ch}),
\end{equation}
where $E$ is the photon energy, $R(E, E_\mathrm{ch})$ is the detector response, $E_\mathrm{ch}$ is the energy channel, and $\tau(t, E)$ is the atmospheric optical depth along the line of sight. The calculation of $\tau(t, E)$ is the same as in \cite{2023ApJS..264....5X}, using the atmospheric model NRLMSIS-2.0 \cite{Emmert2021NRLMSIS}. To model the occultation data, we also assume that the background during the occultation is flat enough so that the background rate for each energy channel can be fit through
\begin{equation}
    b(t, \mathrm{ch}) = \mathrm{e}^{b_0 + b_1 t}.
\end{equation}
Then the logarithmic likelihood for the occultation data can be calculated as
\begin{equation}
    \log \mathcal{L}(A, \gamma) = \sum D \log(s+b) - \Delta t (s+b),
\end{equation}
where $D$ is the count recorded in channel $E_\mathrm{ch}$ and time bin $\Delta t$. The summation is over the 20-200 keV energy channels of the two GRDs and time bins with bin width of 0.5 s from $T_0+1800$ to $T_0+1865$ s. The data modeling method used here is quite similar to \cite{2023ApJS..264....5X}, but in this case our goal is to fit the parameters of the source.

\begin{figure}
\centering
\includegraphics[width=0.6\textwidth]{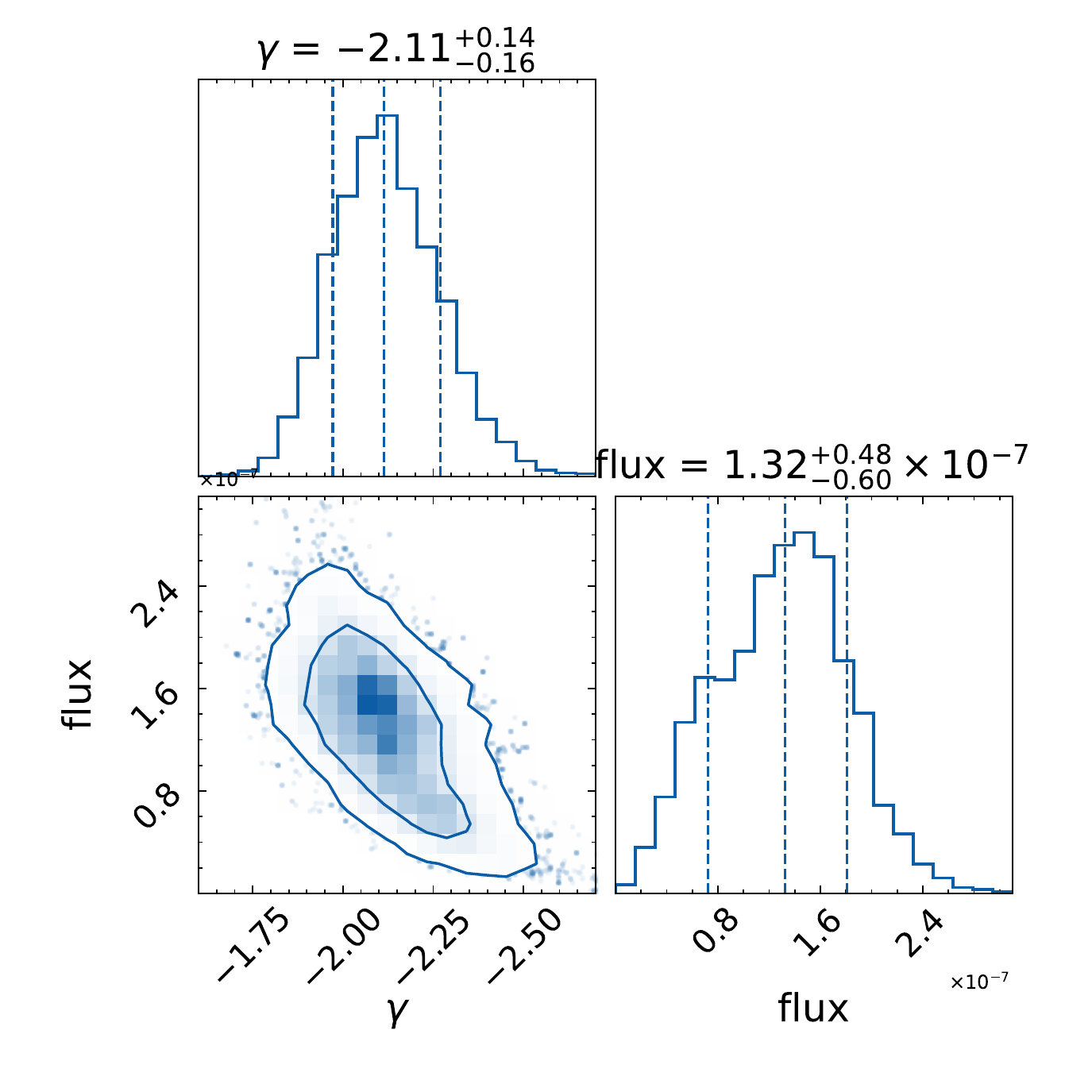}
\caption {Posterior distribution of parameters of interest obtained from occultation analysis.}
\label{fig:OccultResult}
\end{figure}

After assigning a flat prior to $A$, $\gamma$, and the background parameters $b_0$ and $b_1$ of each energy channel, we run the MCMC sampler \cite{Foreman-Mackey2013EmceeMCMC  } to sample these parameters. Integrating the background parameters, we obtain the joint distribution of $\gamma$ and 20-200 keV energy flux, as shown in Figure \ref{fig:OccultResult}. The $68\%$ credible region for 20-200 keV energy flux is $1.32_{-0.60}^{+0.48}\times 10^{-7}$ erg/cm$^2$/s and that for $\gamma$ is $-2.11_{-0.16}^{+0.14}$.

\begin{figure}[htbp]
    \centering  
    \subfigure{
    \includegraphics[width=0.49\textwidth]{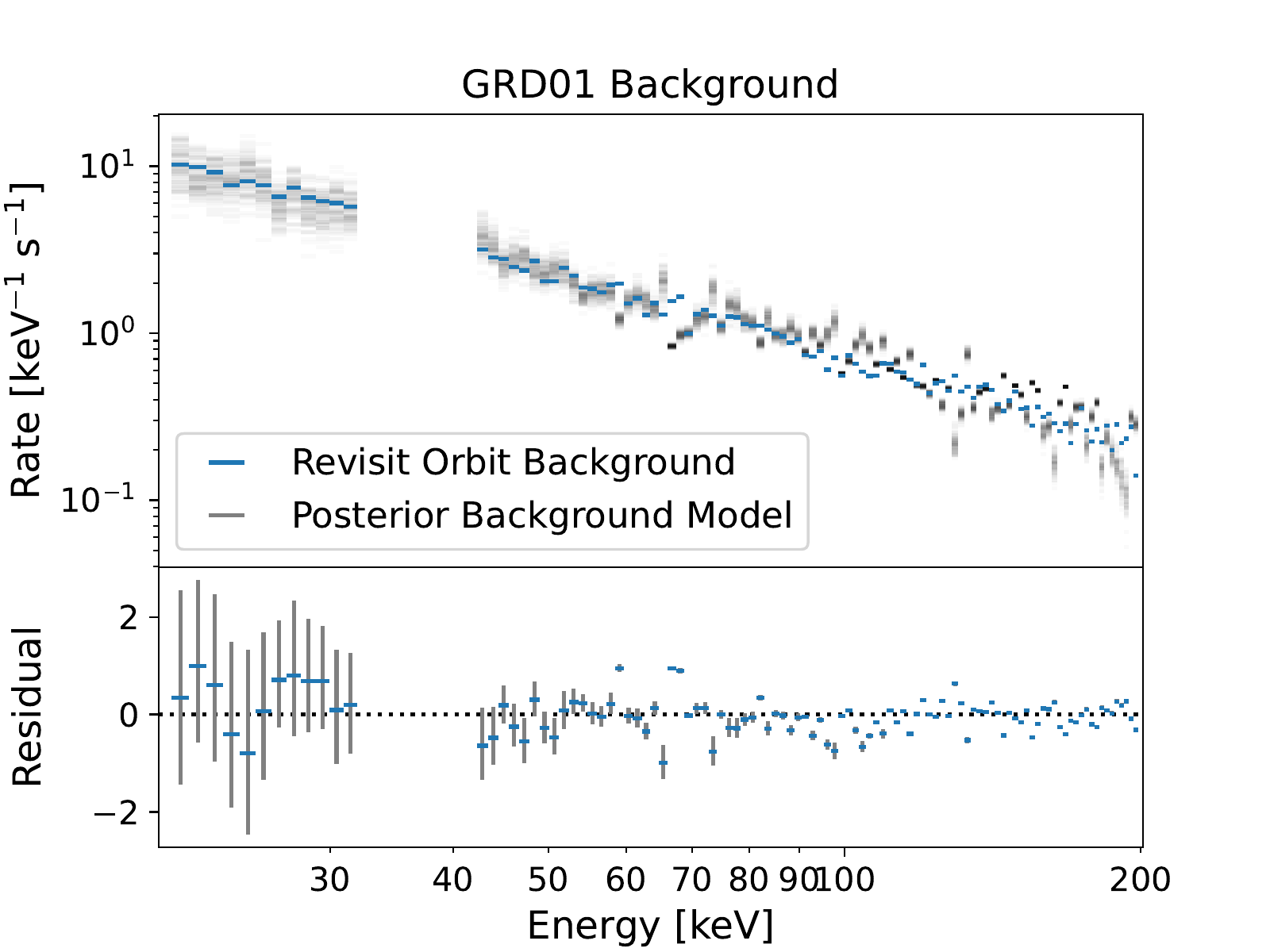}}
    \subfigure{
    \includegraphics[width=0.49\textwidth]{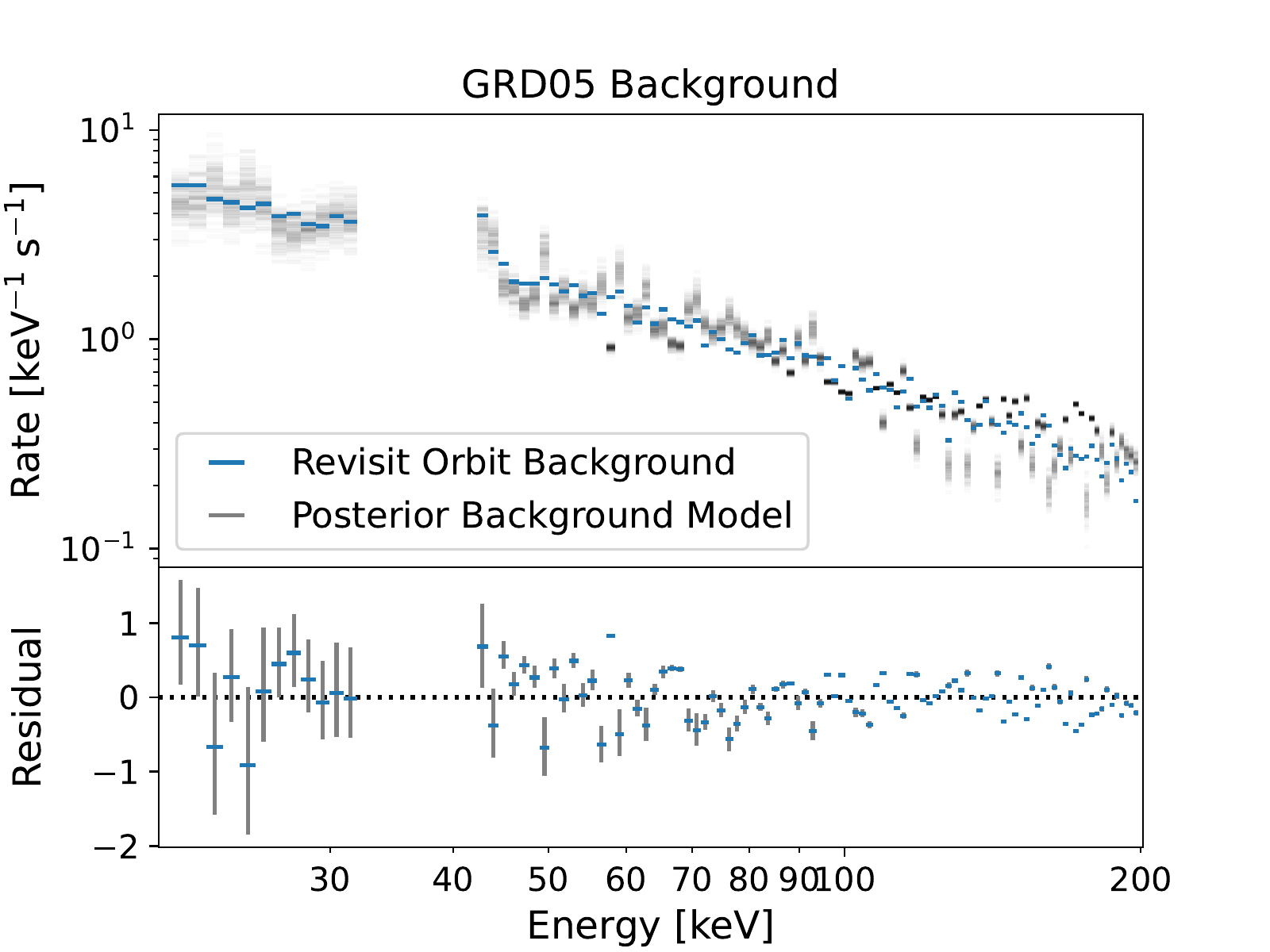}}
    \caption{Comparison of the spectral background estimate from the revisit orbit and that extracted from the posterior model. The lower end, mid, and upper end of the residuals are the difference of the observed background spectra between 15.9\%, 50\%, and 84.1\% of the model values, respectively. The data gap between 32 keV and 42 keV is due to the existence of absorption edges, where the data cannot be perfectly describe by the response matrix currently. The data within this energy range is ignored in the spectral analysis.}
    \label{fig:OccultBack}
\end{figure}

The occultation analysis can also be used to verify the background estimate from the revisit orbit. The background models can be extracted from the posterior distribution and checked if they are consistent with the one estimated from the revisit orbit. We find that these two estimates are consistent during the time interval from $T_0+1800$ s to $T_0+1865$ s, and the comparison is shown in Figure \ref{fig:OccultBack}. 

\subsubsection*{Analysis of the break in light curve}
\begin{figure}[htbp]
\centering  
\includegraphics[width=0.48\textwidth]{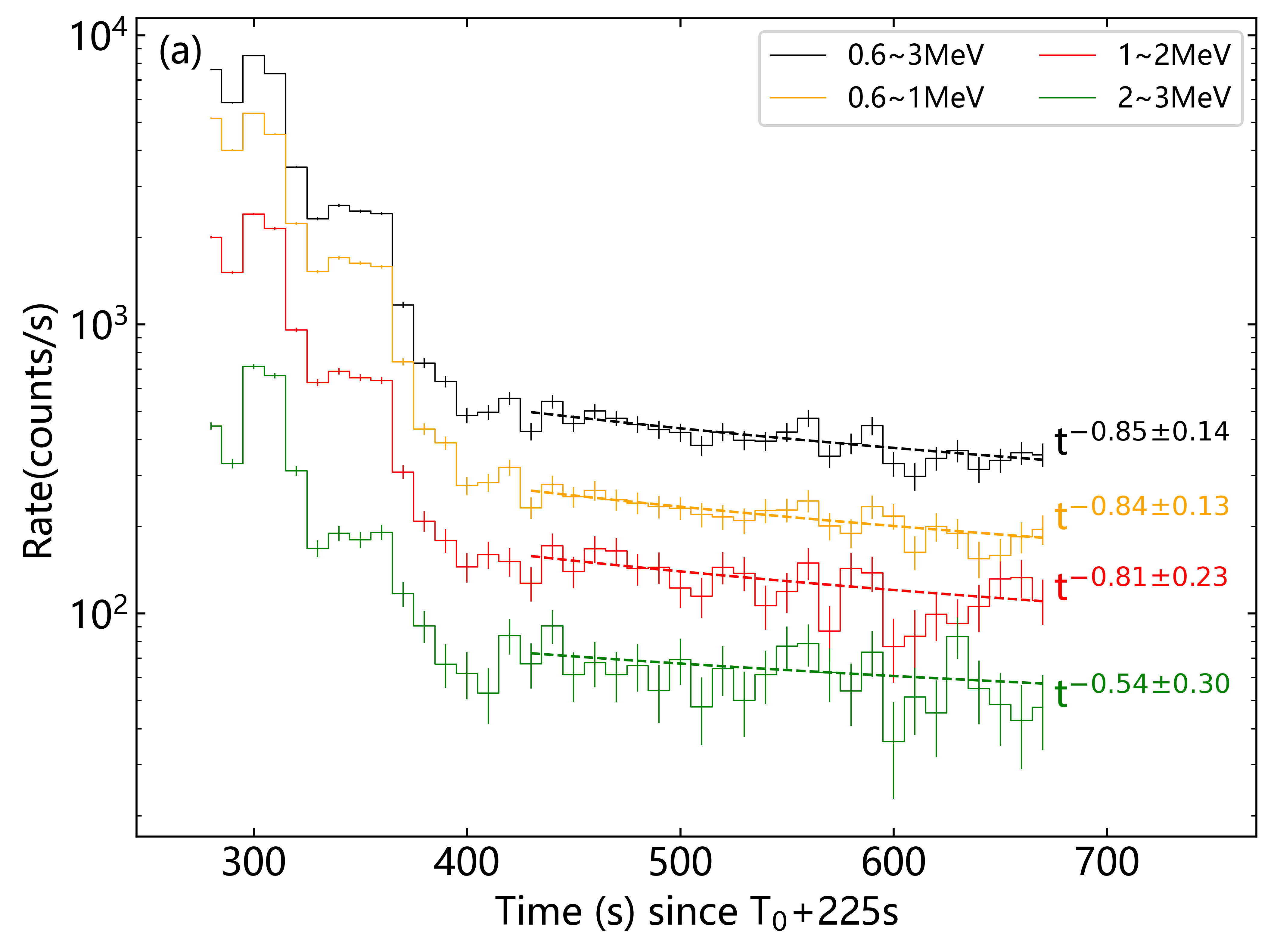}
\includegraphics[width=0.48\textwidth]{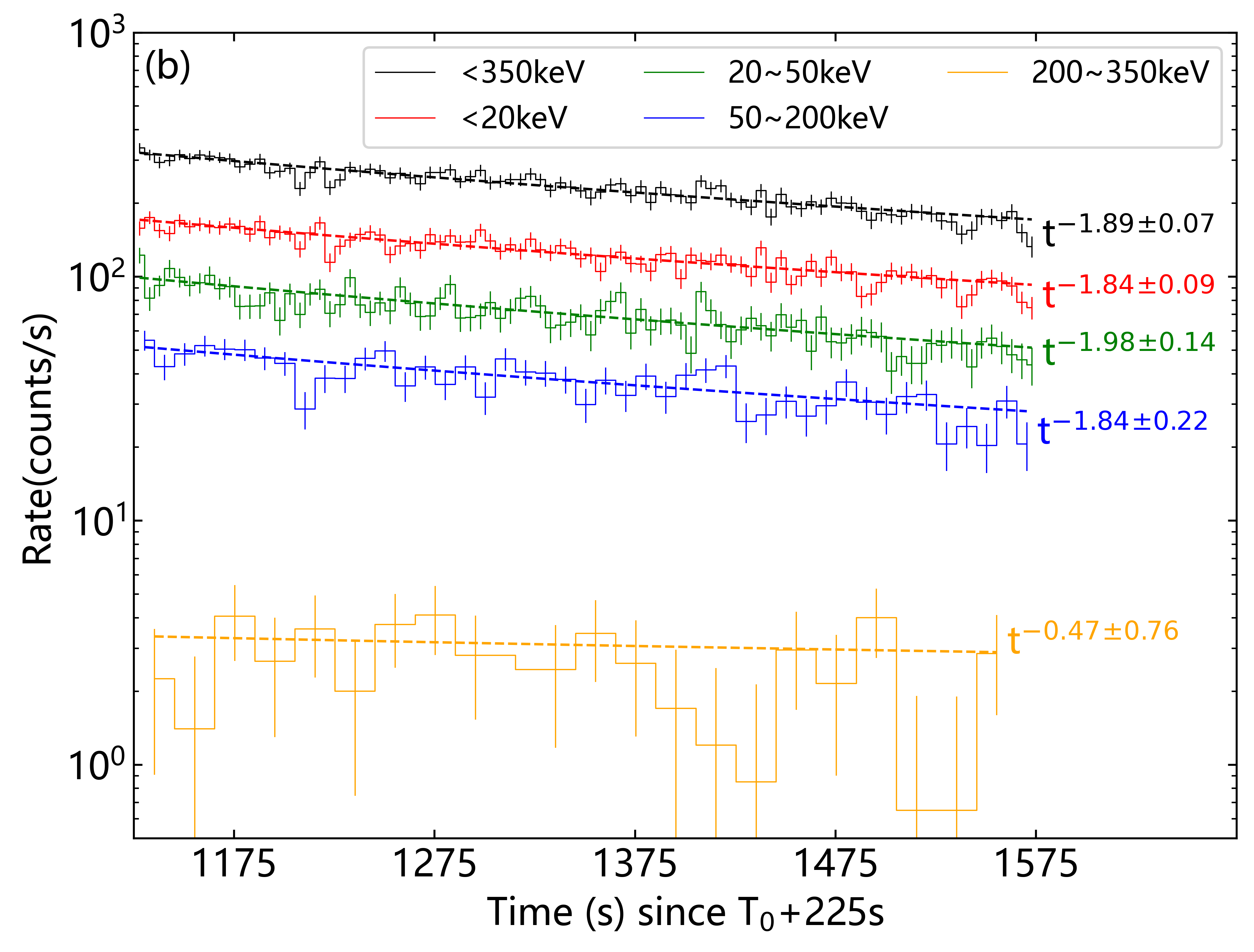}
\caption{Fitting of net light curves of \textit{Insight}-HXMT and GECAM-C during the afterglow period. Panel (a) shows the fitting result of \textit{Insight}-HXMT from $T_0$+540 s to $T_0$+900 s; panel (b) shows the fitting result of GECAM-C from $T_0$+1350 s to $T_0$+1750 s.}
\label{fig:netLC_Fit}
\end{figure}

\begin{figure}
\centering
\includegraphics[width=0.9\textwidth]{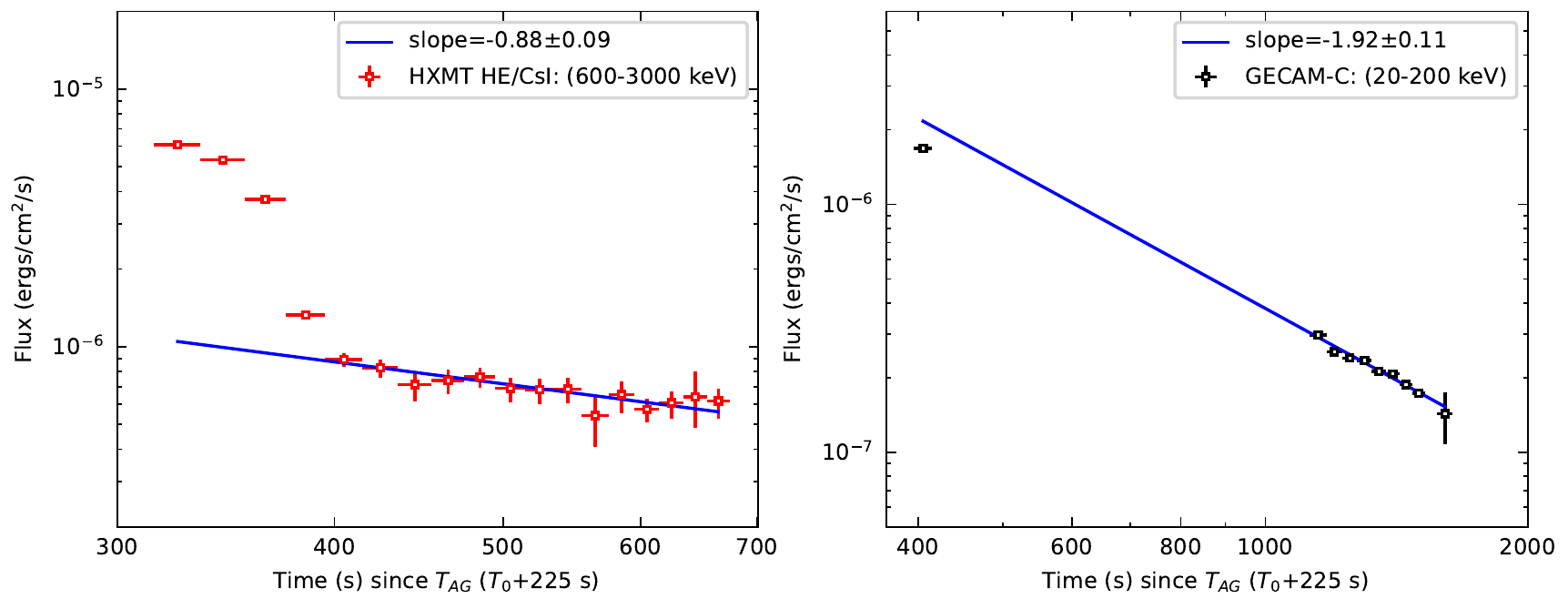}
\caption {Fitting of the early afterglow light curve measured by \insight~HE/CsI (600-3000 keV) and GECAM-C (20-200 keV) respectively.}
\label{fig:raw_flux}
\end{figure}

\begin{figure}
\centering
\includegraphics[width=0.5\textwidth]{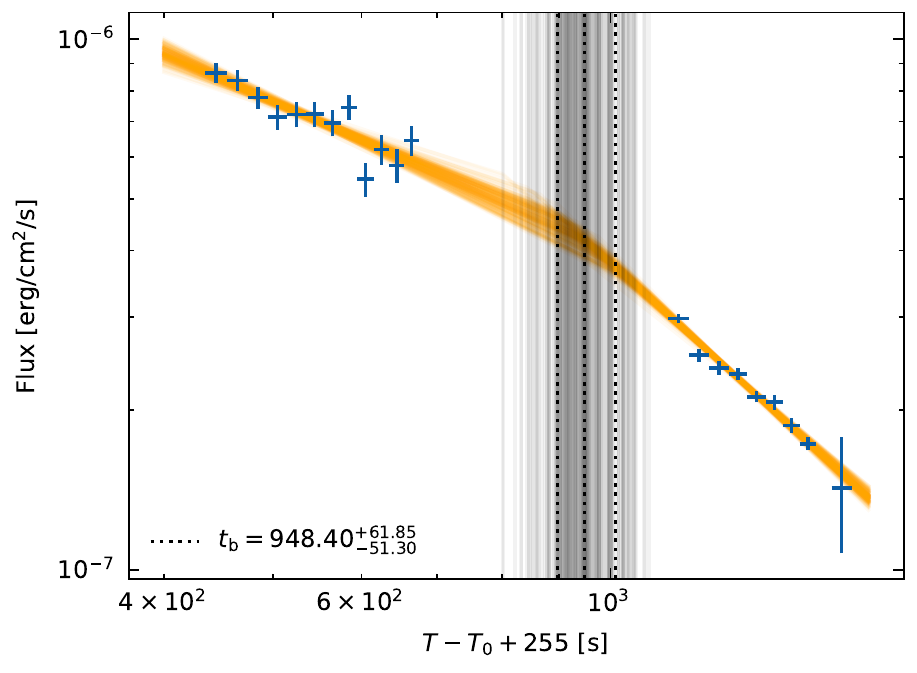}
\includegraphics[width=0.45\textwidth]{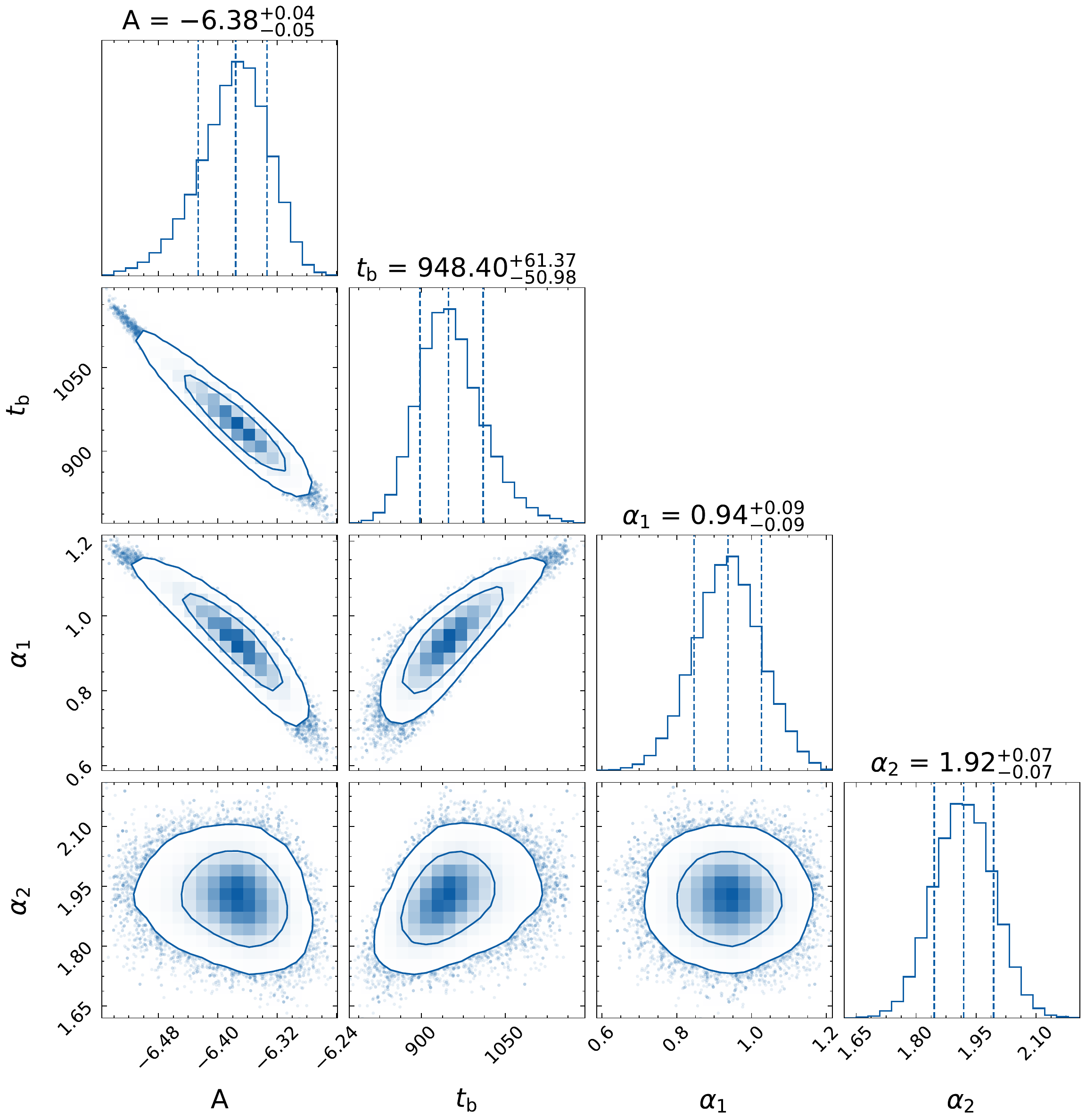}
\caption {Fitting of the early afterglow light curve measured by \insight ~and GECAM-C. The orange lines show posterior distribution of BPL, and the gray vertical lines show the posterior distribution of break time.}
\label{fig:breakfit}
\end{figure}

To derive the energy flux of the afterglow, a detailed time-resolved spectrum analysis is done with a power-law model. We generated 18 time-resolved spectra for \insight~HE/CsI from $T_0$+540 s to $T_0$+900 s with the same time interval 20 s and 8 time-resolved spectra for GECAM-C/GRD01 $\&$ GRD05  from $T_0$+1350 s to $T_0$+1750 s with the same time interval 50 s. By fitting the time-resolved spectrum, the power-law model photon index could be given and the energy flux from 20 to 200 keV of GECAM-C/GRD could also be calculated, as well as for HE/CsI from 600 to 3000 keV. We notice that the slope of energy flux of HE/CsI is not consistent with GECAM-C/GRD (see Fig.\ref{fig:FlareToAfterglow} panel (b)), implying that a break appears between $T_0$+900 s and $T_0$+1350 s. Two photon index in the early afterglow was mentioned in main text (about -1.66 in the high energy band and -2.12 in the low energy band), thus we assume HE/CsI photon index as -1.66 (frozen) and fit time-resolved spectrum from T$_0$+640 s to T$_0$+900 s again to extrapolate the energy flux in 20-200 keV. It is worth noting that the flux error is somewhat underestimated due to the frozen of photon index. The ratio of energy flux in 20-200 keV with power-law model in two photon index could be calculated as a conversion factor when the two power-law intersection confirmed. Through the conversion factor, we could estimate the HE/CsI energy flux in 20-200 keV with photon index -2.12. Therefore, the broken power-law function was utilized to fit the energy flux which combined the HE/CsI estimating energy flux and GECAM-C energy flux in 20-200 keV energy band. Figure \ref{fig:breakfit} shows that a energy flux break appeared about 950 s with 6.8 sigma significance given by likelihood ratio test. The energy flux slope was 0.94 and 1.92 in pre-jet-break and post-jet-break, which was consistent with fitting energy flux in its own energy band showing in Figure\ref{fig:raw_flux} and fitting net light-curve flux directly showing in Figure\ref{fig:netLC_Fit} within error.

As GECAM-C maintains its inertial orientation until the source is blocked by the Earth (see Figure \ref{fig:IncidentAngle}), the incident angle of GRB to GECAM-C detectors keep the same value, then the trend of the net light curve should be consistent with the trend of flux. 
Although \textit{Insight}-HXMT was making scanning observations during the burst, the change of the zenith angle is very small, which has little effect on the effective area in the high energy band. The fitting results can be seen in Figure \ref{fig:netLC_Fit}. The slopes of the light curves of different energy bands are well consistent, which strongly suggests that the spectrum of afterglow has no change in the both periods. Especially, the slope of GECAM-C light curves is about -2, the typical decay slope after jet break, and this implies the jet break occurred before $T_0$+1350s. 

\begin{figure}
\centering
\includegraphics[width=0.45\textwidth]{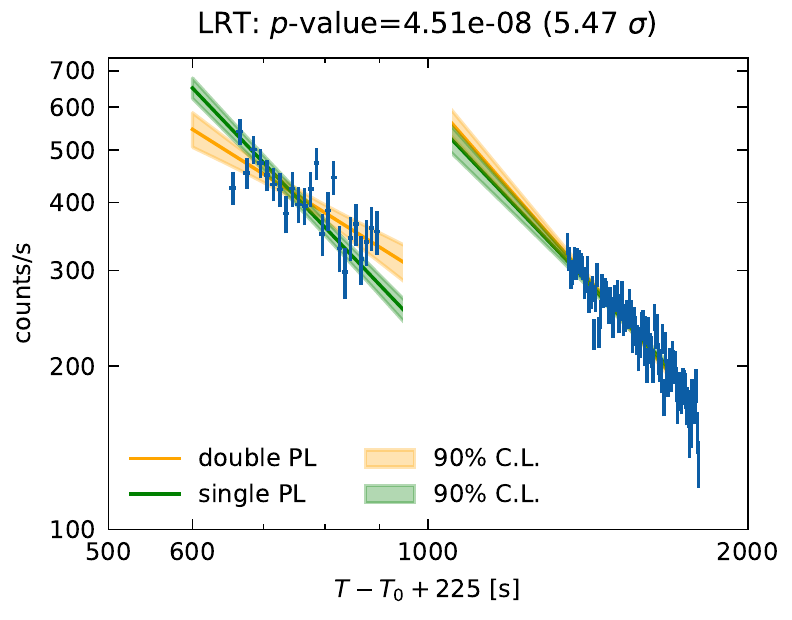}
\includegraphics[width=0.45\textwidth]{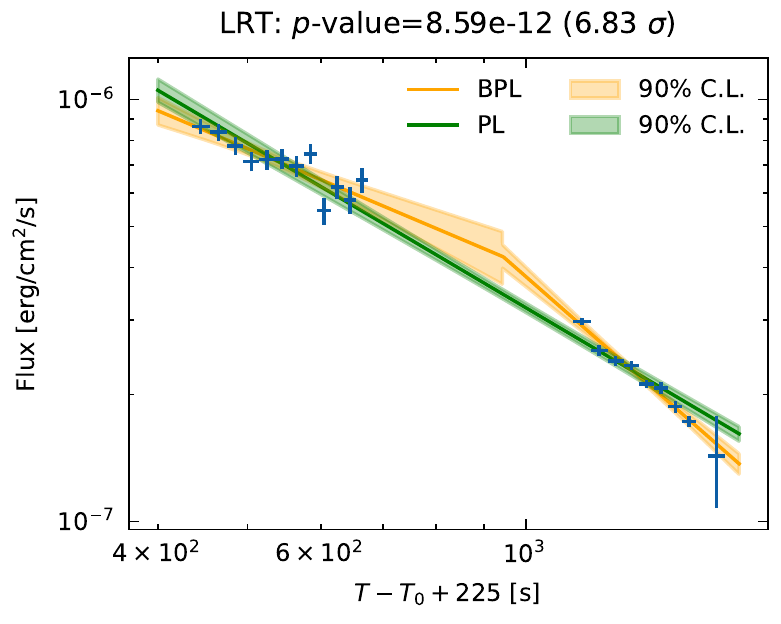}
\caption {Estimate significance of break in net light-curve (left panel) and energy flux (right panel) respectively.}
\label{fig:breaksignificance}
\end{figure}

To investigate whether the flux of the two time segments have the same decay slopes, we performed two different hypothesis tests, both assumes the spectral shape of the afterglow remains the same. This assumption is reasonable as the spectral index appears not to vary with time (see Figure \ref{fig:FlareToAfterglow}). We first test if the 600-3000 keV net light curve obtained from HE/CsI and 20-200 keV net light curve obtained from GECAM/GRD share the same decay index. The net light curve of these two segments is first fit with two decay indices and two normalization factor using Gaussian likelihood, and the likelihood ratio is computed against that with only one decay index. The resulting $p$-value is $4.51 \times 10^{-8}$, corresponding to a 5.47 $\sigma$ significance. Again, the 20-200 keV flux extrapolated from 600-3000 keV data of HE/CsI together with 20-200 keV flux of GECAM/GRD is tested. The flux is fit with a broken power-law (BPL),
\begin{equation}
\begin{split}
        f_{\rm BPL}(t) = \left \{
         \begin{array}{ll}
           A_{\rm BPL} (t / t_{\rm break}) ^ {-\alpha_1}, & t < t_{\rm break} \\
           A_{\rm BPL} (t / t_{\rm break}) ^ {-\alpha_2}, &  t > t_{\rm break} \\
         \end{array}
       \right.
\end{split}
\end{equation}
and a simple power-law (PL),
\begin{equation}
    f_{\rm PL}(t) = A_{\rm PL} (t / t_{\rm ref}) ^ {-\alpha}.
\end{equation}
The free parameters of BPL is amplitude $A_{\rm BPL}$, break time $t_{\rm break}$, and two slope indices $\alpha_1$ and $\alpha_2$. The free parameters of PL is amplitude $A_{\rm PL}$ and slope $\alpha$, while $t_{\rm ref}$ is fixed to the best-fit value of $t_{\rm break}$ in BPL. The likelihood ratio test between BPL and PL yields a $p$-value of $8.59 \times 10^{-12}$, which is equivalent to a significance of 6.83 $\sigma$. These results are shown in Figure \ref{fig:breaksignificance}, and indicate a break is required to explain the flux evolution.

\clearpage

\clearpage

\bibliographystyleMethod{naturemag}
\bibliographyMethod{method}

\end{document}

%% file: authors.tex
%(\textit{Insight}-HXMT \& GECAM Collaboration) 
\author{Zheng-Hua An$^{1}$, 
S.~Antier$^{29}$,
Xing-Zi Bi$^{24}$,
Qing-Cui Bu$^{2}$, 
Ce Cai$^{3}$, 
Xue-Lei Cao$^{1}$, 
Anna-Elisa Camisasca$^{4}$, 
Zhi Chang$^{1}$, 
Gang Chen$^{1}$, 
Li Chen$^{5}$
% ...
}

%% file: affiliations.tex
\begin{affiliations}
    \item Key Laboratory of Particle Astrophysics, Institute of High Energy Physics, Chinese Academy of Sciences, 19B Yuquan Road, Beijing 100049, China  
    \item Institut f\"ur Astronomie und Astrophysik, Kepler Center for Astro and Particle Physics, Eberhard Karls Universit\"at, Sand 1, 72076 T\"ubingen, Germany 
    \item College of Physics and Hebei Key Laboratory of Photophysics Research and Application, Hebei Normal University, Shijiazhuang, Hebei 050024, China  
    \item Department of Physics and Earth Science, University of Ferrara, Via Saragat 1, 44122 Ferrara, Italy
    \item Department of Astronomy, Beijing Normal University, Beijing 100875, China  
    \item CAS Key Laboratory for Research in Galaxies and Cosmology, Department of Astronomy, University of Science and Technology of China, Hefei 230026, China  
    \item Department of Astronomy, Tsinghua University, Beijing 100084, China
    \item Southwest Jiaotong University, Chengdu, 610092, China 
    \item INAF – Osservatorio di Astrofisica e Scienza dello Spazio di Bologna, Via Piero Gobetti 101, 40129 Bologna, Italy
    \item University of Chinese Academy of Sciences, Chinese Academy of Sciences, Beijing 100049, China  
    \item College of physics Sciences Technology, Hebei University, No. 180 Wusi Dong Road, Lian Chi District, Baoding 071002, Hebei, China   
    \item INFN – Sezione di Ferrara, Via Saragat 1, 44122 Ferrara, Italy
    \item Physics and Space Science College, China West Normal University  
    \item Yunnan Observatories, Chinese Academy of Sciences, 650011 Kunming, Yunnan Province, People’s Republic of China      
    \item Key Laboratory of Space Astronomy and Technology, National Astronomical Observatories, Chinese Academy of Sciences, Beijing 100012, China   
    \item School of Astronomy and Space Science, Nanjing University, Nanjing 210093, China  
    \item School of Physics and Optoelectronics, Xiangtan University, Yuhu District, Xiangtan, Hunan, 411105, China   
    \item Guizhou Provincial Key Laboratory of Radio Astronomy and Data Processing, Guizhou Normal University, Guiyang 550001, People’s Republic of China   
    \item Institute of Astrophysics, Central China Normal University, Wuhan 430079, China 
    \item Nevada Center for Astrophysics and Department of Physics and Astronomy, University of Nevada Las Vegas, NV 89154, USA  
    \item Qufu Normal University, Qufu 273165, China  
    \item College of Science, China Three Gorges University, Yichang 443002, China 
    \item Research Center for Intelligent Computing Platforms, Zhejiang Laboratory, Hangzhou 311100, China   
    \item Innovation Academy for Microsatellites, Chinese Academy of Sciences, Shanghai, 201210, People’s Republic of China
    \item National Space Science Center, Chinese Academy of Sciences, Beijing, 100190, People’s Republic of China
    \item Center for Astronomical Mega-Science, Chinese Academy of Sciences, 20A Datun Road, Chaoyang District, 100012 Beijing, People’s Republic of China
    \item Key Laboratory for the Structure and Evolution of Celestial Objects, Chinese Academy of Sciences, 650011 Kunming, People’s Republic of China
    \item Department of Engineering Physics, Tsinghua University, Beijing 100084, People’s Republic of China
    \item Artemis, Observatoire de la C\^ote d'Azur, Universit\'e C\^ote d'Azur, Boulevard de l'Observatoire, 06304 Nice, France
    \item IJCLab, Univ Paris-Saclay, CNRS/IN2P3, Orsay, France
    \item IRAP, Universit\'e de Toulouse, CNRS, UPS, 14 Avenue Edouard Belin, F-31400 Toulouse, France
    \item Universit\'e Paul Sabatier Toulouse III, Universit\'e de Toulouse, 118 Route de Narbonne, 31400 Toulouse, France
    \item Hessian Research Cluster ELEMENTS, Giersch Science Center, Max-von-Laue-Stra{\ss}e 12, Goethe University Frankfurt, Campus Riedberg, 60438 Frankfurt am Main, Germany
    \item School of Physics and Astronomy, University of Minnesota, Minneapolis, Minnesota 55455, USA
\end{affiliations}